# Systems Interoperability Types: A Tertiary Study

RITA S. P. MACIEL, Federal University of Bahia, Brazil
PEDRO H. VALLE, Federal University of Juiz de Fora, Brazil
KÉCIA S. SANTOS, Federal University of Bahia, Brazil
ELISA Y. NAKAGAWA, University of São Paulo, Brazil

Interoperability has been a focus of attention over at least four decades, with the emergence of several interoperability types (or levels), diverse models, frameworks, and solutions, also as a result of a continuous effort from different domains. The current heterogeneity in technologies such as blockchain, IoT and new application domains such as Industry 4.0 brings not only new interaction possibilities but also challenges for interoperability. Moreover, confusion and ambiguity in the current understanding of interoperability types exist, hampering stakeholders' communication and decision making. This work presents an updated panorama of software-intensive systems interoperability with particular attention to its types. For this, we conducted a tertiary study that scrutinized 37 secondary studies published from 2012 to 2023, from which we found 36 interoperability types associated with 117 different definitions, besides 13 interoperability models and six frameworks in various domains. This panorama reveals that the concern with interoperability has migrated from technical to social-technical issues going beyond the software systems' boundary and still requiring solving many open issues. We also address the urgent actions and also potential research opportunities to leverage interoperability as a multidisciplinary research field to achieve low-coupled, cost-effective, and interoperable systems.

## 1 INTRODUCTION

Software-intensive systems (SIS) have increasingly resulted from the combination of multiple heterogeneous systems that interoperate suitably and transparently among them to provide complex functionalities, attending users and business needs. In short, SIS are composed of individual systems that need to exchange information to fulfill specific purposes, and their operations depend sometimes on the interactions among heterogeneous, distributed, independent, and large-scale systems [6, 16]. These systems have supported large infrastructures and application domains, such as Industry 4.0, Health 4.0, smart cities, smart farms, and transportation, just to mention a few. For these systems, interoperability is a key factor that assures their existence (i.e., they would not exist without such interoperability). In turn, interoperability refers to the ability of two or more systems or components to exchange information and use it

Authors' addresses: Rita S. P. Maciel, rita.suzana@ufba.br, Federal University of Bahia, Brazil; Pedro H. Valle, Federal University of Juiz de Fora, Brazil; Kécia S. Santos, Federal University of Bahia, Brazil; Elisa Y. Nakagawa, University of São Paulo, Brazil.

for a given purpose [43]. Interoperability also refers to the process of communication among systems or components without generating a technological dependency among them [95]. Moreover, interoperability is not only about systems or network integration but also about other dimensions, such as social [18] and legal interoperability [30], going nowadays beyond the boundary of software systems. Hence, interoperability can be considered a multidimensional concept comprising diverse perspectives coming from different research communities and application domains.

Interoperability has been a challenge since monolith systems started to communicate with others, and distributed systems started to become popular. From middleware, such as CORBA (Common Object Request Broker Architecture)[1] that enabled communication among systems written in different languages and running on different platforms and hardware, to newer solutions, such as API (Application Programming Interface) and web services, a variety of interoperability types and solutions has emerged to deal with the heterogeneity of systems, including systems from different companies and built with diverse technologies, languages, platforms, standards, communication protocols, and even deployed in diverse locations like cloud and mobile devices. Efforts to mitigate the challenges of interoperability include specific solutions for a given domain (e.g., interoperability standards for Industry 4.0 [19]) or theoretical contributions (e.g., an analysis of pragmatic interoperability [7]).

To address the complicated issues associated with interoperability and better understand it, interoperability has been broken down over the years into various types (often referred to as interoperability levels). Some well-known types are *technical*, *syntactic*, and *semantic*. Each type deals with specific barriers to be overcome to make possible systems communication. The increasing number of interoperability types hampers a comprehensive view of those similar or different types, those related among them, and even a wider acceptation of their definitions/understandings. Hence, various classifications of interoperability types (also referred to as interoperability models) emerged, such as LCIM (Levels of Conceptual Interoperability Model) [86] and LISI (Levels of Information Systems Interoperability) [22]. Additionally, interoperability frameworks, such as EIF (European Interoperability Framework) [30], that usually encompass interoperability models, principles, guidelines, and even reference architectures also exist. Although a considerable number of initiatives and solutions to mitigate interoperability problems are available, interoperability is still a big challenge for software projects, particularly for those contemporary, large, and critical domains.

Regarding the three closely related works, they are outdated ([33] published in 2007) or address specific classes of systems (e.g., context-awareness systems [63]) or a given scenario (a comparative study of interoperability assessment models [52]). To the best of our knowledge, an updated and comprehensive panorama of interoperability types (or levels) is still missing. At the same time, several new types have emerged, while others have become outdated, leading to confusion and ambiguity in the current understanding of the interoperability types. In turn, interoperability types has usually guided the development of models such as LCIM and LISI and frameworks such as EIF as in the past, but the lack of a common and updated understanding about them can difficult future efforts. In summary, the main real-world problem to be addressed in this work is the existing barrier due to the misunderstanding about interoperability types that has hampered interoperability solutions and research and, ultimately, has led to extra effort and cost to build SIS. In this scenario, this paper yields six main contributions:

- We offer an **updated, original, and comprehensive panorama of interoperability types** that was rigorously defined through the conduction of a tertiary study, which scrutinized 37 literature reviews, including systematic literature reviews (SLR) and systematic mappings (SM). In turn, a tertiary study can systematically gather evidence and synthesize data and information from secondary studies in a specific area [50, 91].

---

[1] https://https://www.corba.org//

- We provide a **classification of the 36 interoperability types** found in our tertiary study and the existing similarities, overlaps, and differences among them. We also show the evolution of these types over the years and the new types that emerged more recently, reflecting the state-of-the-art scenario of the field.
- We collect **13 interoperability models (six conceptual models and seven interoperability assessment models (IAM))**, when they emerged, their evolution over the years, the existing relations among them, and their interoperability types (or levels). In short, conceptual models encompass interoperability types and relationships among them that are usually hierarchical, while IAM include interoperability types, barriers, achievement levels, and means to measure and evolve interoperability [76]. These models are the all existing ones, as we systematically scrutinized the literature when conduction the tertiary study.
- We gathered **six interoperability frameworks**, which refer to a more complete way (compared to interoperability models and IAM) to deal with interoperability issues and achieve different purposes. These seven frameworks refer to possibly all existing, as we also systematically and carefully looked for them during the conduction of our study.
- We found the **11 main domains that have concerned with interoperability**. We also discovered **various categories of interoperability solutions (including ontologies, platforms, and reference architectures)** proposed for these different domains.
- Based on our tertiary study, we present **eight main findings and the open issues associated with each finding** as well. We also offer **nine future actions and potential research opportunities to be somehow urgently performed**.

Besides improving the communication among of stakeholders (who are interested in SIS interoperability issues), this work foresees implications for researchers and practitioners:

- **Implications for researchers**: Considering that SIS interoperability is still challenging, many opportunities for research exist. With the emergence of new interoperability types, there is a need for further investigation of new interoperability models and frameworks, new relationships among existing types and new ones, and exploration of new methods to evaluate interoperability solutions. Additionally, given the multidisciplinary nature of interoperability, researchers need to examine how social-technical issues (for instance, cultural interoperability) impact interoperability beyond technical aspects.
- **Implications for practitioners**: By providing a clear panorama of all interoperability types, this work can support industry decision-making regarding the design, implementation, and evaluation of interoperability solutions (e.g., interoperability standards, platforms (tools and services), and ontologies). Furthermore, this work can guide practitioners in identifying specific interoperability types relevant to their industry projects and domains, creating interoperable SIS tailored to their particular needs.

It is worth highlighting that our previous work (a symposium paper) offered a preliminary view of systems interoperability focused on its types [78]. This current work differs from that by answering different research questions, covering a more extensive period, discussing the results more deeply, summarizing the main findings and related open issues, and offering urgent research topics for future research.

This work is organized as follows: Section 2 summarizes other tertiary studies in software engineering and the related works that investigated interoperability from a broader view. Next, Section 3 describes the research method used in our work. The planning and execution phases are detailed, including the research questions, search strategy, inclusion and exclusion criteria, and quality assessment criteria; finally, it lists the studies selected in our tertiary study.

Section 4 shows the results of this study; particularly, it presents an overview of studies in Section 4.1 and details the interoperability types, their definitions, similarities, overlapping, and their classification illustrated through a graphical representation in Section 4.2. Next, Section 4.3 offers the interoperability models and frameworks and depicts the interoperability types (or levels) they encompass, as well as their evolution over the years. Section 4.4 addresses the domains that have concerned with interoperability, e.g., Industry 4.0 and health, and also diverse interoperability solutions found for these domains. Section 5 provides a discussion on the results achieved; more specifically, it presents the main findings and open issues in Section 5.1 and the future actions and potential research opportunities that can further support SIS interoperability in Section 5.2. Besides, Section 5.3 presents the threats to the validity of this work and the means used to mitigate them. Finally, Section 6 concludes this work.

## 2 RELATED WORK

To check our work's originality, we first looked for existing tertiary studies regarding interoperability, but none was found except our previous conference paper. Following, to find the benefits of the research method adopted in our study (i.e., tertiary study), we also searched for studies in April 2022 in ACM DL[2], IEEE Xplore[3], Springer[4], ScienceDirect[5], and DBLP[6] using the keywords "*tertiary study*", "*tertiary studies*", and "*tertiary review*". We found 51 studies[7], most of them focused on software engineering. In summary, from the first tertiary study published in 2010 [49], the quantity of studies has increased gradually; for instance, while 2011 to 2013 had three studies each year, 2021 had nine. More than half part of the studies (28 of 51) was published in journals (the preferred one is Information and Software Technology (with 15) followed by Journal of Systems and Software (5)), evidencing that tertiary studies have their value. Most studies considered secondary studies in a 10-year time range and addressed the state of the art of a specific research topic, such as agile software engineering [40, 96], blockchain [46, 99], software reuse [11], software product line [72], and machine learning. Besides summarizing secondary studies, some presented taxonomies [100] or categorizations [5], also offering perspectives for future work.

Various works have addressed interoperability types differently, not necessarily through tertiary studies. For instance, some studies focused on specific interoperability types or models, exploring their features, benefits, and limitations [38]. Others investigated interoperability of specific software systems domains [79],[71], analyzing the challenges and opportunities related to different types of interactions. Regarding other related works, [63] analyzed systems interoperability considering new and complex software systems that support people's daily activities, and [52] presented a comparative study of IAM. Both studies concluded interoperability conceptualization should be revisited. Furthermore, Ford et al. [33] presented a survey of interoperability assessment and found 64 interoperability types, from which operational, organizational, and technical were the most recurrent. The authors also claimed that interoperability types and their definitions are the bases for proposing interoperability solutions and assessment models. In 2007 when this work was published, the authors concluded interoperability is a research field that should be promoted and refined. In turn, comparing this work with ours, it is outdated, and a more detailed comparison is presented in Section 5, after presenting our results.

---

[2]https://dl.acm.org
[3]https://ieeexplore.ieee.org
[4]https://www.springer.com/gp
[5]https://www.sciencedirect.com
[6]http://dblp.org
[7]The list of tertiary studies is available in https://bit.ly/3Zb87v9

Finally, we can observe tertiary studies make it possible to draw comprehensive views of a specific field (and when a good number of secondary studies already exist) and help researchers identify knowledge gaps in the field. Hence, we adopted tertiary study as the research method of this work.

## 3 RESEARCH METHOD

We performed the three-phase process — planning, execution, and results synthesis [50] — as presented in Figure 1. Sections 3.1 and 3.2 summarize the planning and execution phases respectively, while Section 4 presents the results.

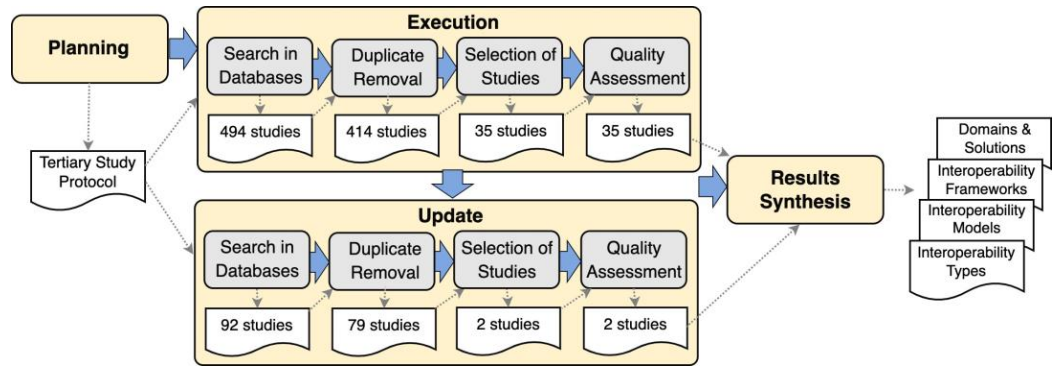

Fig. 1. Research method

### 3.1 Planning

The main elements of the tertiary study protocol are the study goal, research questions (RQs), search string, selected databases, and studies selection criteria. Concerning the goal, we used GQM (*Goal Question Metric*) to define it. The goal is to **analyze** secondary studies of SIS interoperability **for the purpose of** identifying the panorama **with respect to** interoperability types, models, frameworks, domains, and solutions **from the point of view of** researchers in **the context of** academic literature. We also defined three RQs, as listed below:

- $RQ_1$: Which interoperability types have been addressed?
  *Rationale:* Several interoperability types exist, but without an updated and complete panorama of them or a compilation of their definitions/understandings or classification.
- $RQ_2$: Which are the existing interoperability models and frameworks?
  *Rationale:* Interoperability models and frameworks can show how interoperability types can be organized and related to each other. Besides, a complete view of these models and frameworks is missing.
- $RQ_3$: Which are the domains and their related solutions for interoperability?
  *Rationale:* Interoperability solutions can support the interoperability achievement in SIS in those domains (i.e., the contexts where interoperability is addressed by the studies).

After several calibrations through pilot studies, the resulting search string was: *("interoperability" OR "interoperabiliting" OR "interoperable") AND ("review of studies" OR "structured review" OR "literature review" OR "literature analysis" OR "in-depth survey" OR "literature survey" OR "meta-analysis" OR "systematic mapping")*. We also selected the most recommended publications databases for conducting literature reviews: Scopus[8], ACM DL, and IEEE Xplore. For the studies selection, we defined two inclusion criteria (IC) and six exclusion criteria (EC):

- $IC_1$: Study addresses interoperability in SIS.
- $IC_2$: Study addresses interoperability types.
- $EC_1$: Study does not address interoperability in SIS.
- $EC_2$: Study does not address interoperability types.
- $EC_3$: Study is not a literature review.
- $EC_4$: The full text of the study is not available.
- $EC_5$: Study is an older version of another study.
- $EC_6$: Study is not written in English.

We also applied the quality assessment of studies due to our tertiary study intends to present a panorama of all secondary studies, and such assessment can drawn an overall view of the quality of these studies. For this, we followed the recommendation of Kitchenham et al. [49] by using DARE-4 criteria[9]; hence, the questions for the quality assessment (QA) were:

- $QA_1$: Were the review's selection criteria (inclusion and exclusion criteria) explicitly defined?
- $QA_2$: Was the literature search likely to have covered all relevant studies?
- $QA_3$: Did the reviewers assess the quality of the included studies?
- $QA_4$: Was the basic studies' information adequately described?

Each study was scored as follows: $QA_1$ (*Yes*: the selection criteria were explicitly defined; *Partly*: the selection criteria were implicit; or *No*: the selection criteria were not defined and were not possible to be inferred); $QA_2$ (*Yes*: the authors either searched in four or more databases and included additional search strategies or identified and referenced all journals addressing the topic of interest; *Partly*: the authors searched in three or four databases with no extra search strategies, or searched a defined but restricted set of journals and conference proceedings; or *No*: the authors searched up to two databases or a highly restricted set of journals); $QA_3$ (*Yes*: the authors explicitly defined quality criteria and applied them to each primary study; *Partly*: quality criteria and application procedure were implicit; or *No*: no explicit quality assessment of primary studies was presented); and $QA_4$ (*Yes*: information about each study was presented; *Partly*: only summarized information was presented; or *No*: no information was presented). Moreover, the numerical scores were *Yes = 1*, *Partly = 0.5*, *No = 0*, or *Unknown* (the information is not specified).

### 3.2 Execution and Update

We searched for studies in the databases in May 2022, following the protocol rigorously. To do that, we configured the search string to each database's search engine and retrieved studies based on titles, keywords, and abstracts. We focused on studies published in the last ten years (2012 to 2022), similar to most tertiary studies in the literature. We retrieved a total of 494 studies from the databases and after duplicate removal, 414 unique studies remained. During the studies selection process, we deeply examined each study, applied the selection criteria, and selected 35 studies that were relevant to answer our RQs. Next, we applied the quality assessment; for that, the two first authors evaluated the studies separately, and the third and fourth ones came together and solved divergences.

As we finished the deep analysis of each study and results summarization in December 2022, we decided to update our tertiary study considering the period of 2022 until February 2023, as illustrated in Figure 1. We identified two new studies published in 2023. Hence, our tertiary studies selected a total of 37 studies published from January 2012 to February 2023.

---

[8] https://www.scopus.com
[9] https://web.archive.org/web/20070918200401/https://www.york.ac.uk/inst/crd/faq4.htm

## 4 RESULTS

This section presents an overview of studies and answer the three RQs.

### 4.1 Overview of Studies

Table 1 summarizes some information of the 37 secondary studies. This table shows the study ID (used along with this work), review type (i.e., SLR (systematic literature review), qSLR (quasi-systematic literature review), SMS (systematic mapping of studies), LR (Literature Review), and MLR (Multivocal Literature Review)), study title, publication year, the number of primary studies analyzed by each secondary study, quality assessment, and references. Regarding the publication venue, we observed that 27 studies were published in journals and 10 in events. Hence, most studies have high quality and are detailed enough to be published in journals; therefore, we believe they are suitable to answer our RQ. In turn, we indirectly analyzed more than 1,480 primary studies (summing only the number of studies declared in the secondary studies).

The last column of Table 1 lists the calculation of the quality assessment[10]. Most studies with lower scores (0.5 to 1.5) refer to LR, which usually does not adopt a systematic and rigorous way to be conducted; on the other hand, most studies with higher scores (2.0 to 4.0) are SMS or SLR that require systematic review processes. We consciously decided to keep all 37 studies for further analysis, i.e., we did not use the quality assessment to exclude studies. This is because the primary purpose of our study was to provide a comprehensive overview of all review studies on interoperability types published in the field. Second, using DARE allowed us a holistic view of the quality of the studies. Hence, by making the scores available, readers can have a notion of their quality.

Following, Sections 4.2, 4.3, and 4.4 answer $RQ_1$, $RQ_2$, and $RQ_3$, respectively.

### 4.2 Types of Interoperability

Regarding $RQ_1$ (*Which interoperability types have been addressed?*), we examined each study to identify the interoperability types explicitly reported in each one and found 36 different types, as listed in Table 2. We can observe that some types have caught more attention than others; for instance, *semantic* interoperability appeared in 30 studies (including seven studies that specifically addressed it), while *network* occurred in just one (S28). We also extracted the definitions (or understandings) of each type that the studies provided. Four secondary studies (S8, S16, S19, and S24) did not offer the definitions or generically present them; hence, we turned to the primary studies considered in these secondary studies. It is worth highlighting the types (in Table 2) are according to those presented by the studies; hence, ones apparently similar (e.g., *business*, *process*, and *business process*) were considered distinct due to their slight different definitions. We concerned to keep the original information from the studies, considering it could be important for further analyses by other interested. We also observed that interoperability types are sometimes referred to as levels, stages, layers, phases, or dimensions and are usually organized through a hierarchical relationship (further detailed in Section 4.3). In the end, we collected a total of 117 definitions[11] distributed to 36 types.

Deeply analyzing each type's definition(s), we gathered similarities, overlaps, and even divergences. Due to this scenario, we believe that a grouping and positioning of the interoperability types may favour understanding them

---

[10]The detailed calculation is available in https://bit.ly/3Zb87v9 .

[11]All definitions of each interoperability type and the raw data extracted from studies and used to answer $RQ_1$ are available at https://bit.ly/3Zb87v9.

Table 1. List of selected studies

| ID | Review Type | Title | Year | # Studies | QA |
|---|---|---|---|---|---|
| S1 | LR | Enterprise information systems' interoperability: Focus on PLM challenges [31] | 2012 | N/A | 0.5 |
| S2 | LR | Research methodology for enterprise interoperability architecture approach [38] | 2013 | N/A | 0.5 |
| S3 | LR | A review of interoperability standards in e-health and imperatives for their adoption in Africa [1] | 2013 | N/A | 0.5 |
| S4 | LR | A review on e-business interoperability frameworks [75] | 2014 | N/A | 0.5 |
| S5 | LR | Generic data models for semantic e-government interoperability: Literature review [77] | 2014 | 7 | 1.0 |
| S6 | SLR | Interoperability evaluation models: A systematic review [76] | 2014 | N/A | 1.0 |
| S7 | SLR | Clinical information modeling processes for semantic interoperability of electronic health records: Systematic review and inductive analysis [61] | 2015 | 36 | 2.0 |
| S8 | SLR | Measuring tool chain interoperability in cyber-physical systems [37] | 2016 | 42 | 2.5 |
| S9 | SLR | Semantic web technologies in cloud computing: a systematic literature review [17] | 2016 | 36 | 2.5 |
| S10 | SLR-SMS | Towards pragmatic interoperability to support collaboration: A systematic review and mapping of the literature [68] | 2016 | 13 | 3.5 |
| S11 | LR | Interoperability and portability approaches in inter-connected clouds: A review [47] | 2017 | 120 | 1.0 |
| S12 | qSLR | Rethinking interoperability in contemporary software systems [63] | 2017 | 17 | 1.0 |
| S13 | SLR | Semantic interoperability for an integrated product development process a systematic literature review [83] | 2017 | 14 | 2.5 |
| S14 | SLR | A systematic review to merge discourses interoperability, integration and cyber-physical systems [36] | 2018 | 48 | 3.5 |
| S15 | LR | Interoperability governance: a definition and insights from case studies in Europe [98] | 2018 | N/A | 0.5 |
| S16 | LR | Semantic interoperability in industry 4.0: Survey of recent developments and outlook [69] | 2018 | 13 | 0.5 |
| S17 | qSLR | A conceptual perspective on interoperability in context-aware software systems [62] | 2019 | 17 | 2.0 |
| S18 | LR | A Review of interoperability standards for Industry 4.0 [19] | 2019 | N/A | 0.5 |
| S19 | SLR | A systematic literature review of interoperability in the green building information modeling life cycle [64] | 2019 | 115 | 1.5 |
| S20 | LR | Exploring semantic interoperability in e-government interoperability frameworks for intra-African collaboration: a systematic review [101] | 2019 | 10 | 2.5 |
| S21 | SLR | Interoperability assessment: A systematic literature review [52] | 2019 | 38 | 1.5 |
| S22 | SMS | IoT Semantic interoperability A systematic mapping study [90] | 2019 | 16 | 3.0 |
| S23 | SMS | Semantic interoperability in IoT: A systematic mapping [81] | 2019 | 32 | 0.5 |
| S24 | SLR | Semantic interoperability methods for smart service systems: A survey [20] | 2019 | 75 | 3.5 |
| S25 | SLR | Ontology-based solutions for interoperability among product lifecycle management systems: A systematic literature review [34] | 2020 | 54 | 4.0 |
| S26 | SLR | Toward semantic IoT load inference attention management for facilitating healthcare and public health collaboration: A survey [55] | 2020 | 15 | 3.0 |
| S27 | LR | Developing a transnational health record framework with level-specific interoperability guidelines based on a related literature review [53] | 2021 | 9 | 1.5 |
| S28 | SLR | A Systematic review on the data interoperability of application layer protocols in industrial IoT [4] | 2021 | 34 | 3.5 |
| S29 | SLR | The fast health interoperability resources (FHIR) standard: Systematic literature review of implementations, applications, challenges and opportunities [10] | 2021 | 19 | 4.0 |
| S30 | LR | A survey on blockchain interoperability: Past, present, and future trends [12] | 2021 | 102 | 0.5 |
| S31 | SLR | Semantic interoperability in health records standards: A systematic literature review [60] | 2022 | 28 | 4.0 |
| S32 | SLR | Interoperability requirements for blockchain-enabled electronic health records in healthcare: A systematic review and open research challenges [73] | 2022 | 176 | 0.5 |
| S33 | LR | Blockchain oracles: State-of-the-art and research directions [32] | 2022 | 280 | 0.5 |
| S34 | MLR | Security and privacy challenges in blockchain interoperability - A multivocal literature review [41] | 2022 | 46 | 1.5 |
| S35 | SLR | Interoperability of heterogeneous health information systems: a systematic literature review [88] | 2023 | 36 | 1.0 |
| S36 | SLR | Blockchain for healthcare management systems: A survey on interoperability and security [92] | 2023 | 21 | 2.5 |
| S37 | LR | Exploring blockchains interoperability: A systematic survey [94] | 2023 | 14 | 0.5 |

better, so we prepared Figure 2. We grouped the types into **technological**, **social-technical**, and **crosscutting**. For this work, we defined that **technological group** refers to those types that deal with the interaction among SIS elements, such as hardware, networks, software platforms, and different systems. In turn, this group was split into **low level** and **high level**. While types in the **low level** group are necessary to overcome heterogeneity in software infrastructures (e.g., *technical*, *objects*, and *device*), types in **high level** meet the demands to support the collaboration among different software elements to achieve SIS goals (e.g., *semantic*, *syntactic*, and *pragmatic*). **Social-technical group** contains types that address the capability of elements (outside the software itself), people, and organizations to collaborate and overcome their differences. This group was split into **individual** and **organization** groups. While the **individual**

Table 2. Types of interoperability addressed by secondary studies

| ID | Interoperability types |
|---|---|
| S1 | Organizational, Semantic, Technical |
| S2 | Business, Data, Enterprise, Process, Service |
| S3 | Organizational, Semantic, Syntactic, Technical |
| S4 | Business, Cloud, Cultural, Data, Ecosystems, Electronic identify, Knowledge, Objects, Organizational, Process, Rules, Semantic, Service, Social networks, Software systems, Syntactic, Technical |
| S5 | Legal, Organizational, Semantic, Technical |
| S6 | Cloud, Cultural, Data, Ecosystems, Electronic identify, Knowledge, Objects, Organizational, Process, Rules, Semantic, Service, Social networks, Software systems, Syntactic, Technical |
| S7 | Semantic |
| S8 | Conceptual, Enterprise, Operational, Organizational, Process, Programmatic, System, Technical |
| S9 | Semantic |
| S10 | Pragmatic, Semantic, Syntactic |
| S11 | Semantic |
| S12 | Semantic, Syntactic, Technical |
| S13 | Organizational, Semantic, Technical |
| S14 | Constructive, Data, Enterprise, Functional, Information, Operational, Process, Programmatic, System, Technical |
| S15 | Legal, Organizational, Semantic, Technical |
| S16 | Data, Enterprise, Information, Semantic, System, Technical |
| S17 | Conceptual, Organizational, Semantic, Syntactic, Technical |
| S18 | Semantic |
| S19 | Business, Data, Organizational, Process, Service, Technical |
| S20 | Organizational, Semantic, Technical |
| S21 | Coalition, Constructive, Legal, Operational, Pragmatic, Semantic, Syntactic, Technical |
| S22 | Semantic |
| S23 | Hardware, Platform, Semantic |
| S24 | Device, Platform, Semantic, Syntactic, |
| S25 | Organizational, Semantic, Technical |
| S26 | Semantic, Syntactic, Technical |
| S27 | Business process, Organizational, Semantic, Syntactic, Technical |
| S28 | Data, Device, Network, Platform, Semantic |
| S29 | Semantic |
| S30 | Blockchain, Legal, Organizational, Semantic, Technical |
| S31 | Semantic |
| S32 | Blockchain, Organizational, Semantic, Syntactic |
| S33 | Blockchain |
| S34 | Blockchain |
| S35 | Dynamic, Conceptual, Functional, Pragmatic, Semantic, Syntactic, Structural, Technical |
| S36 | Blockchain |
| S37 | Blockchain, Semantic |

**group** focuses on addressing issues related to people, the **organization group** focuses on addressing issues related to organizations. *Cultural* is an example of a type of social-technical individual interoperability, and *enterprise* is a social-technical organization type. Finally, the **crosscutting group** impacts both technological and social-technical aspects of SIS interactions; *legal* was then placed in this group.

Another analysis was regarding the stability of the definitions, i.e., how much the definitions of a given interoperability type have somehow been kept the same in different studies. For this, we carefully analyzed all definitions found for each type, making it possible to group it into four groups — **stable**, **ongoing**, **initial understanding**, and **misunderstanding** — colored in green, blue, orange, and red, respectively, in Figure 2. The more aligned the definitions of the given type, the more stable that type is. In the context of this work, **stable understanding** refers to types with minimal disagreement in their definitions and present in several studies over the years. For instance, *semantic* interoperability is addressed in 30 studies published from 2012 to 2023 (all period considered in our study). **Ongoing understanding** encompasses types that are not well-defined regarding both the type name and its definition. This group have more overlapping than discrepancies. Coincidentally, all five types associated with business process alignment (in blue in Figure 2) are in this group, possibly because they deal with interoperability in different domains, such as industry (S19), health (S27), and enterprise (S2 and S4). **Initial understanding** refers to those types addressed in few studies (one or two studies), and no relevant disagreement in their definitions was observed. The type in **misunderstanding** presented discrepant definitions, such as *data* with six non-aligned definitions coming from seven different studies, and *functional* and *programmatic* with two distinct definitions each from two studies.

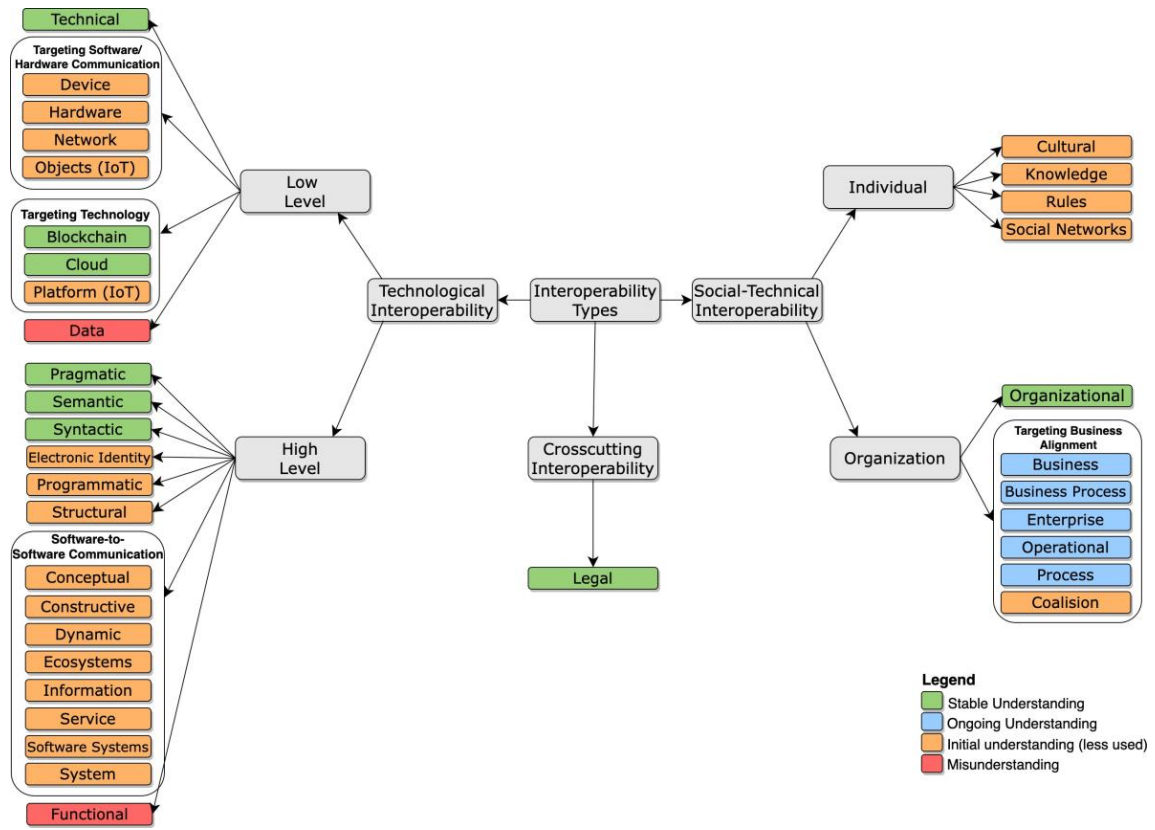

Fig. 2. Classification of interoperability types

In summary, few types (i.e., eight types) have a stable definition. The majority (21 types) have an initial understanding, followed by ongoing understanding (five types), and two was classified as misunderstanding. Hence, while considerable attention has been given to several different types, the concern has not been unifying the definitions/understandings. Following, we provide a discussion of each group of interoperability types, highlighting the most relevant issues observed in each group and type.

Concerning the **technological** and **low level group**, nine types are in these group, as shown in Figure 2. All these types address software communication across heterogeneity of hardware devices, network, and systems infrastructures. In particular, *technical* interoperability is a concern in most studies (20 of 37) and enables the communication among software systems despite the heterogeneity of underlying communication infrastructures. Its definitions often comprise various solutions, such as standardization of hardware and software interfaces (as in S5, S21, and S25). While *technical* interoperability is domain-independent, other types, such as *device*, *hardware*, *network*, *objects*, and *platform*, appear more in specific technological domain, e.g., IoT-based devices and solutions (as in S23 and S24). Additionally, while *network* and *hardware* keep similarities with *technical* interoperability (i.e., communication between hardware and software infrastructure), *platform* has similarities to *syntactic* interoperability (i.e., to overcome the heterogeneity of communication protocols). *Cloud* interoperability treats heterogeneity among cloud providers and cloud layers (as in S4 and S6). *Blockchain* interoperability appeared in six recent studies (from 2021) and surprisingly already presents a common definition considering these studies. This type refers to the ability of distinct and heterogeneous blockchain systems to work together seamlessly, enabling information exchange/sharing and access across different blockchain networks without a centralized authority, each representing a distributed data ledger (S30, S32, S33, S34, S36, and S37). S30 still points out that *blockchain* interoperability refers to *technical* and *semantic* interoperability. *Data* interoperability is the one that presents significant divergent definitions found in seven studies (S2, S4, S6, S14, S16, S19, and S28). When defined for IoT context, it refers to the ability to deal with heterogeneity in the acquisition and use of data across different platforms and software systems; in other words, it seems similar to *technical* interoperability. S28 defines *data* interoperability as one composed of *device*, *network*, *platform*, and *semantic* interoperability. When considering specific domains, such as cyber-physical systems (in S14) and constructive industry (in S19), the purpose is to make data accessible despite different representation strategies, such as languages, syntax, and data models, i.e., it is similar to *syntactic* interoperability.

Considering **technological** and **high level group**, *semantic*, *syntactic* and *pragmatic* have stability in their definitions, other 11 types have initial understanding, and one presents misunderstanding definition. *Semantic* interoperability appears in most studies (30 of 37) and provides communication without information ambiguity and in an accurate way without human intervention, so software systems need to agree with a common information representation. Although *semantic* is usually treated as domain-independent interoperability, it has drawn attention from specific domains, e.g., Industry 4.0 (S16, S18, and S28), health (S3, S7, S26, S27, S29, S31, S32, and S35), and IoT (S22, S23, and S24). In particular, IoT-focused studies (S22 and S23) present slightly different definitions, in which the heterogeneity of data sources (image, text, and video) captured by sensors and delivered to applications may be solved through ontologies; hence, it is related to a problem of *semantic* interoperability. Besides, there is a trend that the heterogeneity and specificity in these domains should be addressed differently, i.e., through low-level interoperability types; so *device* (as in S24 and S28), *network* (S28), and *platform* (S23, S24, and S28) are mentioned. Similarly to *semantic* interoperability, another type with long-term investigation is *syntactic* found in 12 studies. In general, it refers to structural and behavioral aspects of message exchange to enable communication among distinct software systems. Another stable type is *pragmatic* that aims the communication by sharing context, intention, and effect of message exchanges (S10, S21, and S35).

Regarding software-to-software communication, we gathered eight types. In short, *ecosystems* refers to the ability of different ecosystems collaborate among them and with independent entities (S4 and S6). *Conceptual* (S8, S17, and S35) refers to the highest interoperability level in SIS, when several aspects to achieve interoperability are aligned (e.g., process, information, restrictions, and contexts). *Service* highlights that interaction among software services should be dynamically performed (S2, S4, S6, and S19). *Information* (S14 and S16), *software systems* (S4 and S6), and *system* (S8, S14, and S16) address systems interaction and effective use of information, while *constructive* (S14 and S21) and *dynamic* (S35) occur when SIS are aware about state changes in the assumptions and limitations they are making over time.

The remaining three types with initial understanding of the **technological** and **high level group** deal with specific interoperability issues: (i) *electronic identity* refers to systems from different domains that collaborate considering automatic authentication and authorization concerns (S4 and S6); (ii) *programmatic* was proposed to attend peculiarities (e.g., activities workflow) of specific domains (S14); and *structural* refer to when multimedia, hypermedia, object-oriented data, and other forms of information are recorded. Finally, *functional* presents a misunderstanding; while S14 defines it as an ability to exchange information reliably, S35 refers to this type as functional requirements delivered by the systems in a consistent and established manner.

The **social-technical group** deals with social-technical aspects related to organizations and people's needs and contains 11 types, as shown previously in Figure 2. The **social-technical** and **individual group** comprises four types with initial understanding and addressed in S4 and S6 only. These types are newer (compared to others) and cope with interoperability beyond the software boundaries: (i) *cultural* (to be used by transnational organizations to deal with different languages, traditions, religions, and ethics while collaborating in the same domain); (ii) *knowledge* (to promote sharing of intellectual assets and repositories); (iii) *rules* (to align and match business and legal rules on organizations' automated transactions); and (iv) *social networks* (to promote the ability of enterprises to use and interconnect with social networks for collaboration purposes).

**Social-technical** and **organization group** comprises seven types. The *organizational* type, a stable type, aims to align business processes to carry out cross-organizational interactions and is a concern of 15 studies (S1, S3, S4, S5, S6, S8, S13, S15, S17, S19, S20, S25, S27, S30, and S32). All five interoperability types with ongoing understanding follow business guidelines to promote interaction among organizations and process alignment (as addressed by nine studies: S2, S4, S6, S8, S14, S16, S19, S21, and S27). In general, these types seem to be the same but with some differences in their definitions and were proposed mainly by domain-specific studies. For instance, *business* is addressed in enterprise domain (S2 and S4), *process* in cyber-physical systems (S8 and S14), *business process* in health (S27), and *enterprise* in Industry 4.0 (S16). Furthermore, *operational* (S8, S14, and S21) is a specific type focused on project's cost and time, and process failures as well. From a broader view, these five types with ongoing understanding and domain-specific could be considered sub-types of *organizational* (which is a domain-independent type).

With respect to **crosscutting group**, it contains only one type (*legal*) found in four studies (S5, S15, S21, and S30). *Legal* interoperability may impact both technological and social-technical aspects of SIS; hence, we considered it as having a crosscutting nature. We also considered *legal* has a stable understanding and concerns the legal aspect of business processes and software functionalities, such as government laws and enterprise policies. In particular, S5 and S15 addressed e-government, evidencing *legal* interoperability can be relevant for this domain.

In summary, the eight stable types (*technical*, *blockchain*, *cloud*, *pragmatic*, *semantic*, *syntactic*, *legal*, and *organizational*) seem to have already a consensus regarding their definition in the literature. Otherwise, most types (21) still have an initial understanding, with few studies addressing them. Moreover, five types (*business*, *business process*, *enterprise*, *operational*, and *process*), all of them directly related to *organizational* interoperability, have ongoing understanding, and *data* and *functional* present misunderstanding, despite the number of studies that mention it.

After deeply analyzing each interoperability type, its definitions, and possible categorization (as previously shown in Figure 2), we considered the publication year of the secondary studies and drew Figure 3. For instance, seven studies were published in 2019 and mentioned *semantic*. The intention of this figure was to provide an overall view of the stability of types and quantity of secondary studies addressing them and, therefore, the interests in those types over the years. The year 2019 had the most studies published (eight studies), as previously listed in Table 1, so this year concentrates many types. At the same time, the three studies published in 2014 also concentrate many types. Additionally, *semantic* was the most recurrent one (in 30 studies) followed by *technical* (20), *organizational* (15), and *syntactic* (12); in particular, they are addressed in most interoperability models and frameworks (e.g., LCIM encompasses *technical*, *syntactic*, *semantic*, and *pragmatic*, while EIF addresses *organizational*) (Section 4.3 further discusses models or frameworks). Besides, *semantic* seems to keep the interest over the years. We also observe that due to the increasing interest in IoT, cloud computing, Industry 4.0, and other domains that require smart systems, *technical* has also drawn attention due to the high heterogeneity of hardware and software infrastructure and the lack of standards for that. As a highlight, *organizational* (with a stable understanding) is the most recurrent among the social-technical ones. In turn, this type has been a concern since enterprises started demanding interoperability among their systems, as found in S1, which addressed interoperability in enterprise information systems. Additionally, the most recurrent types in the secondary studies also present stable understanding (i.e., *organizational*, *semantic*, *syntactic*, and *technical*). Finally, *blockchain* seems to be a new trend having six studies associated since 2021.

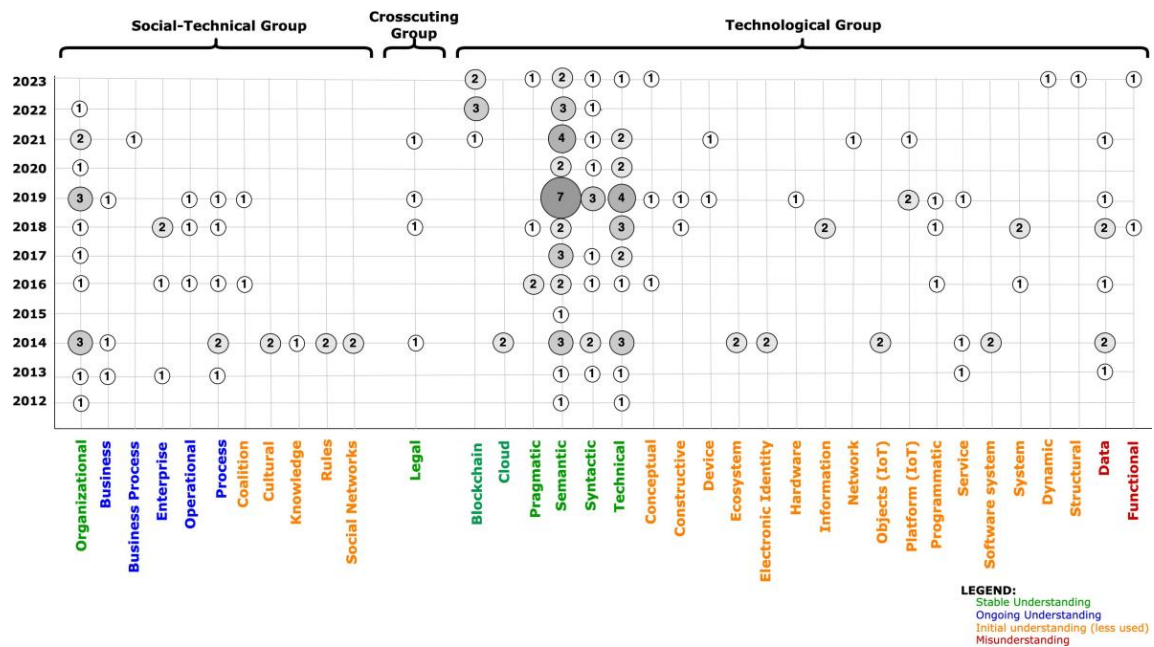

Fig. 3. Distribution of interoperability types (according to the secondary studies where they were found)

In summary, the diversity of interoperability types from various studies reveals the interest and concern of the community in the field of systems interoperability. Many definitions or understandings of interoperability types exist with their specificities, similarities, some overlapping, and even divergences. Moreover, the continuous emergence of new types over the years to solve interoperability issues coming from new technologies (e.g., cloud-based or blockchain) or new domains (e.g., Industry 4.0) indicates that several open issues still need to be solved.

### 4.3 Interoperability Models and Frameworks

Concerning RQ$_2$ (*Which are the existing interoperability models and frameworks?*), we looked for existing models and frameworks (which usually include interoperability models) addressed in the secondary studies. Table 3 shows the 13 different interoperability models (six conceptual models and seven IAM) and six different frameworks found in 12 (of 37) studies. We believe these models and frameworks are the most relevant because the secondary studies contemplated them. Still, we also believe there are others out of the scope of these studies.

Table 3. Interoperability models (conceptual models and interoperability assessment model (IAM)) and frameworks

| ID | Conceptual Model | IAM | Framework |
|---|---|---|---|
| S1 | | | AIF, EIF, INTEROP* |
| S4 | | | EIF, IDEAS, GridWise, INTEROP |
| S5 | | | EIF |
| S6 | MCISI, SoIM | EIMM, LISI, OIM | |
| S8 | LCI, LCIM, MCISI, NMI, SoIM, SoSI | LISI, OIAM, OIM | |
| S10 | LCIM | | |
| S14 | LCI, NMI, SoSI | LISI, OIAM, OIM | AIF |
| S17 | LCIM | LISI, OIM | AIF, INTEROP |
| S20 | LCIM | | |
| S21 | LCIM, SoIM | DIAM, LISI, MMEI, ULSSIMM | EIF, INTEROP |
| S25 | LCIM, LCI | | |
| S30 | | | BIF |

**AIF** (ATHENA Interoperability Framework), **DIAM** (Disaster Interoperability Assessment Model), **EIF** (European Interoperability Framework), **EIMM** (Enterprise Interoperability Maturity Model), **GridWise** (GridWise Interoperability Context-Setting Framework), **IDEAS** (Interoperability Development for Enterprise Application and Software Framework), **INTEROP** (Enterprise Interoperability Framework), **LCI** (Layers of Coalition Interoperability), **LCIM** (Levels of Conceptual Interoperability Model), **LISI** (Levels of Information Systems Interoperability), **MCISI** (Military Communications and Information Systems Interoperability), **MMEI** (Maturity Model for Enterprise Interoperability), **NMI** (NATO C3 Technical Architecture Reference Model for Interoperability), **OIAM** (Organizational Interoperability Agility Model), **OIM** (Organizational Interoperability Maturity Model), **SoIM** (Spectrum of Interoperability Model), **SoSI** (System of Systems Interoperability), **ULSSIMM** (Ultra Large Scale Systems Interoperability Maturity Model), **BIF** (Blockchain Interoperability Framework).
* This framework was originally named EIF (Enterprise Interoperability Framework), but we referred to it as INTEROP in this work to avoid confusion with EIF (European Interoperability Framework).

Interoperability models organize interoperability types and can be a conceptual model or an IAM. A **conceptual model** presents types and relationships among them that are usually hierarchical. An **IAM** encompasses interoperability types, barriers, achievement levels (stages, phases, or dimensions that are commonly associated with one interoperability type or with desired attributes), and means to measure and evolve interoperability [76]. An **interoperability framework** not only organizes the types but also addresses different aspects, including assumptions, concepts, values, practices, software solutions, interoperability achievement guidelines, reference architectures, and others. Hence, frameworks refer to a more complete way (compared to interoperability models) to deal with interoperability issues and achieve different purposes.

We observed there are still misunderstandings concerning the models and frameworks. For example, a study classified LCIM as an IAM, but it is a conceptual model [87]. i-Score was also classified as IAM, while it is indeed a method for interoperability assessment [37] and, therefore, we did not consider it in our study. Another example is e-GIF (found in S20), which refers to any frameworks for e-government domain, then removed from our analysis. Other studies presented misunderstandings regarding the interoperability types contained in models and frameworks. Hence, we assumed that the secondary studies served us to identify relevant models and frameworks. To complement our analysis, we examined primary studies and other external materials, such as technical reports, websites, books, and others.

Figure 4 presents a panorama of the interoperability models (conceptual models and IAM) and frameworks distributed over the years when they were first delivered. Appendix A of this work summarizes the interoperability types or levels of each model and framework. Some models and frameworks were updated: LCIM (in 2005), NMI (NATO C3 Technical Architecture Reference Model for Interoperability) (2007), and EIF (2017), and some served as a basis to derive others. We also observed that some models or frameworks were initiatives of standardization organizations (e.g., IEEE (Institute of Electrical and Electronics Engineers), DTMF (Distributed Management Task Force), and ISO (International Organization for Standardization)) and research institutions like SEI (Software Engineering Institute). Others are from research projects or international calls such as H2020[12] or even government interests. In particular, US military forces have been continually interested in interoperability issues and introduced the three first interoperability models: SoIM (Spectrum of Interoperability Model) in 1980 [51], MCISI (Military Communications and Information Systems Interoperability) in 1996 [3], and LISI in 1998 [21]. Moreover, NATO (North Atlantic Treaty Organization) introduced NMI in 2003 with an interest in promoting interoperability among systems from different nations [66].

---
[12]http://www.ec.europa.eu/horizon2020

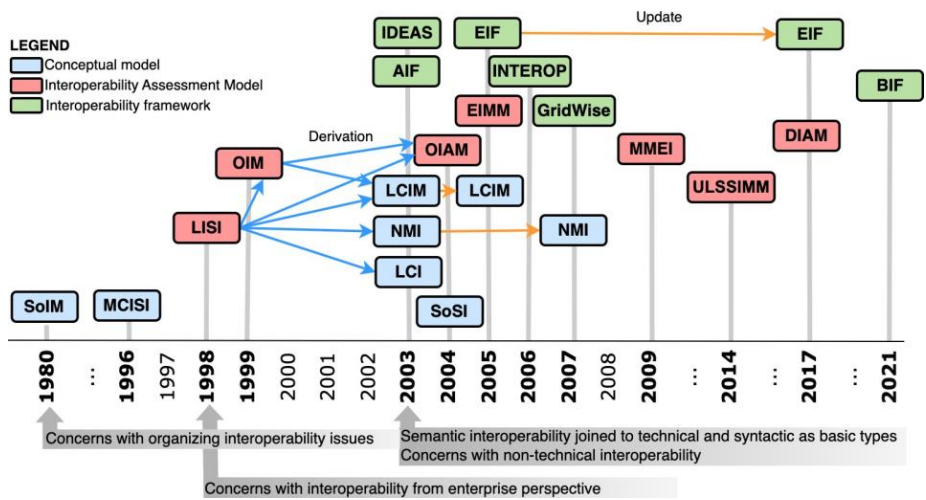

Fig. 4. Interoperability models and frameworks

Backing to the beginning of the 1980s, distinct interoperability issues began to be organized. In the case of **SoIM** and **MCISI**, technical and syntactic aspects of interoperability were taken into account for the military domain. At the end of the 1990s, **LISI** introduced a new proposal based on levels to organize interoperability issues. Moreover, LISI can be considered the first IAM and the first that contemplated enterprise issues as interoperability concerns; hence, it has contributed as a reference to derive or influence other models (e.g., LCI, LCIM, and OIM (Organizational Interoperability Maturity Model)). We can also observe that both MCISI and LISI focused on information systems, probably due to the need to interoperate this class of systems at that time.

In more detail, **LISI** focuses on assessing the interoperability maturity in information systems. It defined five interoperability levels [21]: (i) level 0 (isolated - interoperability in a manual environment); (ii) level 1 (connected - interoperability in a peer-to-peer environment); (iii) level 2 (functional - interoperability in a distributed environment); (iv) level 3 (domain - interoperability in an integrated environment including domain data models and procedures); and (v) level 4 (enterprise - interoperability in a universal environment including enterprise data models and procedures). These levels are not explicitly related to interoperability types, but LISI named these levels and defined the desired characteristics for each level. LISI also defines four attributes within each level (summarized by the acronym PAID (Procedures, Applications, Infrastructure, and Data)) to encompass the range of interoperability considerations. In summary, LISI offers a matrix with the five levels in the rows and the four attributes in the columns. Each matrix' cell describes the abilities or capabilities needed to achieve a specified level of interoperability for a specific attribute.

Still, at the end of the 1990s and derived from LISI, **OIM** emerged as another IAM as an initiative of DST[13] (Australian Defense Science and Technology Organization) and aimed to assess the ability of organizations to interoperate among them considering human-activity aspects [26]. In more detail, OIM presents five levels [26]: (i) independent (no interaction); (ii) ad-hoc (ad-hoc arrangements); (iii) collaborative (shared goals, roles and responsibilities); (iv) integrated (common understanding and preparedness); and (v) unified (shared knowledge base across the systems). OIM also presents four attributes to assess interoperability among organizations [26]: preparedness, understanding, command style, and ethos. Based on OIM, **OIAM** (Organizational Interoperability Agility Model) appeared to address agility attributes (e.g., open and dynamic) as assessment items [48]. It is worth highlighting that OIM, OIAM, **EIMM** [9], and **MMEI** (Maturity Model for Enterprise Interoperability) [35] are domain-independent models, while the other two — **ULSSIMM** (Ultra Large Scale Systems Interoperability Maturity Model) [74] and **DIAM** (Disaster Interoperability Assessment Model) for disaster response management systems [54] — are domain-specific ones. We can observe that besides offering interoperability categorization, these IAM present attributes to interoperability measurement in general. Indeed, such an assessment can determine the strengths and weaknesses of an entity (or system) in terms of interoperability. These models usually offer means (e.g., methodologies or guidelines) to identify the current interoperability level (in which an entity is placed) and achieve the desired level. Hence, IAM show that interoperability, more than a non-functional requirement to be implemented in software systems, is a continuous issue that organizations should pursue.

---
[13]https://www.dst.defence.gov.au

Backing to the beginning of the 2000s and derived from LISI and/or OIM, three interoperability models emerged: LCI, LCIM, and NMI. **LCI** went beyond by dealing with interoperability issues not explicitly addressed in LISI; LCI introduced the concept of coalition interoperability that aims the collaboration of several organizations [84]. LCI presented two layers (technical and organizational), also referred to as layers of coalition interoperability. The technical layer comprises four levels (physical, protocol, data/object model, and information), and the organizational layer contains four levels (aligned procedures, aligned operations, harmonized strategies/doctrines, and political objectives). Between both layers, the knowledge/awareness level appears as intermediate level. Hence, LCI is relevant by first coping with social-technical interoperability issues, particularly through the organizational layer.

Another relevant model is **LCIM**, the most recurrent one in the secondary studies. This model classifies the interoperability issues into seven *conceptual* levels (no interoperability, technical, syntactic, semantic, pragmatic, dynamic, and conceptual). It is worth highlighting that its updated version [89] mapped each level explicitly to an interoperability type. LCIM has influenced many interoperability solutions and other models and frameworks despite its simplicity. For instance, some have adopted the strategy to associate a level to an interoperability type (e.g., EIF, GridWise, INTEROP, AIF, and ULSSIMM), and others have adopted semantic interoperability concerns (e.g., EIF, e-GiF, and ULSSIMM). The LCIM highest levels (i.e., *pragmatic*, *dynamic*, and *conceptual*) focus on diverse core requirements to assure systems collaborations. Otherwise, **NMI** focused specifically on *technical* interoperability and defined four degrees considering manual and automated activities to achieve interoperability among systems: (i) unstructured data exchange (human-readable text); (ii) structured data exchange (human and manual interpretable data); (iii) seamless sharing of data (common and automated data model); and (iv) seamless sharing of information (universal and automated data processing). In its updated version in 2007, NMI kept the degrees but aligned to the LISI levels. Another model introduced at that time was **SoSI**, proposed by SEI. This model is the first that explicitly addresses the interoperability in SoS (Systems-of-Systems), a class of systems with unique characteristics mainly regarding the managerial and operational independence of constituent systems [28]. SoSI considers four interoperability types (technical, programmatic, constructive, and operational), each one associated with a given level.

**Interoperability frameworks** go beyond conceptual models and IAM as they combine several elements to overcome interoperability challenges. Initiatives worldwide exist, such as EIF in Europe and Gridwise in the US. These initiatives are destined for different domains, including enterprises (AIF, INTEROP, and IDEAS), public services (EIF), and electric power (GridWise), and have covered different interoperability types. In particular, **EIF** can be considered one of the most complete and well-known frameworks. EIF was proposed in the context of an H2020 project and provided several strategies and guidance for developing and updating the nations' interoperability frameworks and policies. EIF contributes to establishing a single digital market by fostering cross-border and cross-sectoral interoperability to deliver European public services. This framework addresses interoperability relationships among public services, between public services and non-public enterprises, and between public services and ordinary citizens. EIF comprises [30]: (i) a set of 12 principles to establish general behaviors on interoperability; (ii) a conceptual model for designing and operating interoperable public services; and (iii) a layered interoperability model that organizes different interoperability aspects. This model is considered a primary element in establishing an interoperability-by-design paradigm to the other EIF elements and has four interoperability layers (legal, organizational, semantic, and technical), integrated public governance as a crosscutting layer, and interoperability governance as a background layer. EIRA[14] (European Interoperability Reference Architecture) is another initiative of the same consortium that, aligned with the EIF, specifies four building blocks for developing interoperable public services. Each block corresponds to one of four interoperability types of EIF layers. As we can see, EIF is a comprehensive framework that, in addition to establishing conceptual aspects, provides guides for interoperability implementation, aiming to facilitate its adoption. As EIF targets the European community, legal interoperability is an essential aspect of being considered.

Another well-known framework is **AIF** (ATHENA Interoperability Framework), a result of ATHENA project[15] that provided solutions to meaningful interoperation among enterprises. AIF contains guidelines to address business needs and technical requirements for interoperability based on a multidisciplinary and model-driven solution approaches in solving interoperability problems. AIF encompasses four types of interoperability achievement (from the lowest to the highest): information/data, services, process, and enterprise/business. Semantic aspects are considered a crosscutting barrier that should be overcome at all levels. This framework is structured into three parts: (i) conceptual integration (which focuses on concepts, meta models, languages, and model relationships and provides a modeling foundation for systematizing several aspects of interoperability); (ii) applicative integration (which focuses on methodologies, standards, and domain models and provides guidelines, principles, and patterns that can be used to solve interoperability issues); and (iii) technical integration (which focuses on developing ICT environments). AIF also provides tools and platform specifications for developing and running enterprise applications and software systems. Among other artifacts, AIF comprises: (i) EIMM (Enterprise Interoperability Maturity model, which defines concerns and maturity levels to determine the current ability of an enterprise to collaborate with external entities and specify the path to improve this ability); and (ii) IIAM (Interoperability Impact Analysis Method, which focuses on the return of investment (ROI) and the impact of the interoperability measures).

Similarly to AIF, another enterprise-oriented framework is **INTEROP**[16], proposed by the INTEROP Network of Excellence [24]. INTEROP presents four concerns (business, data, process, and service), three barriers (conceptual, technological, and organizational), and three approaches (federated, unified, and integrated) aiming to identify the barriers in the early phases before implementing the interoperability solutions. **IDEAS** is another framework with an important contribution by highlighting that interoperability among enterprises could be achieved considering different types (application, business, communication, data, and knowledge) [102]. Finally, the U.S. Department of

---

[14] https://joinup.ec.europa.eu/collection/european-interoperability-reference-architecture-eira
[15] https://cordis.europa.eu/project/id/507849
[16] Originally named EIF (Enterprise Interoperability Framework), but referred to as INTEROP in this work to avoid misunderstanding with EIF (European Interoperability Framework).

Energy proposed **GridWise** in 2007 to enable interoperability among various entities that interact with the electric power system [97].

A very recent framework is **BIF** (Blockchain Interoperability Framework), published in 2021 [12], that classifies solutions that connect different blockchain systems, enabling a more comprehensive interoperability analysis. BIF classifies interoperability solutions into public connectors, hybrid connectors, and blockchain of blockchains. As a highlight, all other studies we found and deal with blockchain interoperability are aligned to this framework [32, 41, 92]. Hence, for a while, BIF can be considered a reference for the blockchain context.

We observed that frameworks found in our work consider mainly those interoperability types that are part of *technological group* and with *stable understanding* (as previously shown in Figure 2), such as *technical*, *syntactic*, and *semantic*. These three types seem to form a basic stack for becoming a software system interoperable. Moreover, models and frameworks often address interoperability types in *social-technical group*, particularly related to organizations, such as *organizational*, *business*, and *enterprise*. Finally, there are many interoperability models and frameworks, but few are widely adopted in the industry, as already stated previously in [36]; at the same time, a model with good impact is LCIM, well-known frameworks are EIF and AIF, and a framework with potential wide adoption is BIF, considering the popularity that blockchain has gained.

### 4.4 Domains and Interoperability Solutions

Concerning RQ$_3$ (*Which are the domains and their related solutions for interoperability?*), we found 12 different domains addressed in 33 (of 37) studies, listed in Table 4. It is worth highlighting these domains refer to those as referenced by the studies. We only joined those that can be clearly considered synonymous, like Industry 4.0 and IIoT (Industrial Internet of Things). While some domains represent a segment of the reality for which software systems are developed, such as health and Industry 4.0, others refer to hardware and software infrastructure for supporting the development and execution of systems (e.g., cloud computing and blockchain). We also collected the interoperability solutions (and their categories) for those domains[17] from the secondary studies. Table 4 lists the category of these solutions (i.e., conceptual model, domain-specific model, framework, ontology, meta-model, platform, and standard) and those interoperability types explicitly addressed by these solutions; therefore, these types are a subset of those presented previously in Table 2.

The **health** domain has drawn attention and appeared in nine studies (S3, S7, S26, S27, S29, S31, S32, S35, and S36). In particular, S26 also addressed IoT and both S32 and S36 coped with blockchain. The ability of health information systems to communicate and work together is crucial for efficient management of health services, public health initiatives, high-quality care for patients, and clinical research. When interoperability is lacking, medical information becomes disorganized, repetitive, and difficult to access, leading to lower quality care and a waste of financial resources [88]. In the health domain, *semantic* interoperability seems to be a major concern, in particular, for EHR (Electronic Health Record) based systems that have used ontologies and web standards (e.g., OWL (Ontology Web Language)). However, no consensus about standards for EHR still exists. *Syntactic* and *technical* interoperability have been treated through the adoption of several standards, for instance, HL7 v3 messaging, FHIR (Fast Healthcare Interoperability Resources), openHR, ISO 13606, and DICOM (Digital Imaging and Communications in Medicine). We also observed that health equipment manufacturers usually adopt one of the proposed standards and implement a mapping to other solutions to achieve interoperability. HIMSS[18] (Healthcare Information Management Systems Society), a global advisor for the transformation of the health ecosystem, continuously works on *technical*, *syntactic*, and *semantic* interoperability.

---

[17]The complete list of interoperability solutions is available at https://bit.ly/3Zb87v9.
[18]https://www.himss.org/

Table 4. Domains, categories of interoperability solutions for these domains, and interoperability types addressed by the solutions.

| Domain | ID | Interoperability Solutions Category | Interoperability Types |
|---|---|---|---|
| Blockchain | S30 | Framework, Protocol, Platform (API and Middleware Layers) | Technical, Semantic, Organizational, Legal, Blockchain |
| | S32 | Standards | Semantic, Syntactic, Organizational, Blockchain |
| | S33 | Platforms (Gateway), Oracles, Cryptocurrency-Directed, Blockchain Engines and Connectors | Blockchain |
| | S34 | Protocols | Blockchain |
| | S36 | Frameworks, Platforms (Gateways and API), Proxies | Blockchain |
| | S37 | Technologies and Protocols | Semantic, Blockchain |
| Cloud Computing | S9 | Domain-specific models, platform (tool and technology), ontology, and standard | Semantic |
| | S11 | Ontology, platform (tool and service), and standard | Semantic |
| Collaborative Systems | S10 | Meta model, platform (tool and service), ontology | Pragmatic |
| Constructive Industry | S19 | Domain-specific model | Business, Data, Process, Service |
| Context-awareness software systems | S17 | Framework | Organizational |
| Cyber-physical Systems | S8 | Models | Conceptual, Data, Enterprise, Functional, Organizational, Operational, Process, Programmatic, Systems, Technical |
| | S14 | Conceptual model | Data, Constructive, Enterprise, Functional, Information, Operational, Process, Programmatic, Systems, Technical |
| E-Government | S5 | Conceptual model, framework, ontology | Semantic |
| | S15 | Framework | Legal, Organizational, Semantic, Technical |
| | S20 | Domain-specific model, framework, standard | Semantic |
| Enterprise | S1 | Platform (tool and service) | Technical |
| | S2 | Framework | Business, Data, Enterprise, Process, Service |
| Health | S3 | Framework (architecture and model) and standard | Organizational, Semantic, Syntactic, Technical |
| | S7 | Domain-specific model, framework, and standard | Semantic |
| | S26 | Ontology and platform (gateway and architecture) | Semantic, Syntactic, Technical |
| | S27 | Framework | Business Process, Organizational, Semantic, Syntactic, Technical |
| | S29 | Standard | Semantic |
| | S31 | Ontology, platform (semantic web technologies), and standards | Semantic |
| | S32 | Standards | Semantic, Syntactic, Organizational, Blockchain |
| | S35 | Standards and Technologies | Technical, Syntactic, Pragmatic, Dynamic, Conceptual, Structural, Functional, Semantic |
| | S36 | Frameworks, Platforms (Gateways and API), Proxies | Blockchain |
| Industry 4.0 | S16 | Domain-specific model, framework, ontology, and reference architecture | Semantic |
| | S18 | Domain-specific model, framework, ontology, and standard | Semantic |
| | S28 | Framework, platform (gateway, tool, middleware, and API), protocols, reference architecture, and standard | Data, Device, Network, Platform, Semantic |
| IoT | S22 | Ontology and platform (API, gateway, and service) | Semantic |
| | S23 | Ontology | Semantic |
| | S24 | Ontology, platform (API and service), and standard | Device, Platform, Semantic, Syntatic |
| Manufactury | S13 | Domain-specific model, framework, meta model, and ontology | Semantic |
| | S25 | Ontology and platform (API, middleware and service) | Semantic |

Another interesting initiative is THR (Transnational Health Record) system framework [53], which considers *technical*, *syntactic*, *semantic*, and *business process* interoperability levels to promote health globally, aiming to achieve digital healthcare without borders. Furthermore, the *business* interoperability deals with workflows to several processes, such as user data registration and accesses. The health domain has also worked to achieve interoperability as a non-functional requirement; particularly, the next challenges is the *organizational* type (S3, S27 and S32). Due to the use of IoT in the medical ecosystems and the heterogeneity in health systems, *technical* and *syntactic* interoperability are still issues to be solved (S26). Using blockchain in the health industry can address the challenges faced in cross-domain EHR solutions, ensuring privacy and scalability (S32 and S36). Over the years, interoperability in the health domain has been a constant concern, given that there are studies published from 2015 (S3) to 2023 (S36); in particular, S35 details and classifies several interoperability solutions for health systems. In summary, the health domain has recognized the significance of interoperability as a cross-cutting factor in healthcare systems (S35). In addition, this domain has considered the impact of various technologies, such as IoT (S26) and blockchain (S32, S35, and S36), on different aspects of its systems.

Interoperability is also a primary non-functional requirement for **enterprise** and **e-government** domains (as found in S1, S2, S5, S15, and S20). According to three studies (S2, S5 and S15), the Internet and its communication protocol have supported the interoperability in these domains, leading to the emergence of e-business and e-government concepts. Companies need to align their business rules, policies, and constraints to cooperate among them, and then interoperability must go beyond communication platforms that host software systems. These domains have then invested in frameworks, such as EIF and AIF, as S20 mentioned. Regarding the interoperability types, *technical*, *syntactic*, and *semantic* are also a concern in these domains. Those types (i.e., *business*, *enterprise*, and *process*) associated with *organizational* interoperability also appeared. Additionally, *legal* seems relevant when dealing with government systems, but this type is still little studied.

**Industry 4.0** addresses the challenges of enabling end-to-end communication among all production-relevant assets on the shop floor and information technologies [14]. The complexity of this domain requires solutions for different interoperability types. S28 details an extensive list of solutions (e.g., protocols, API, gateways, and middleware), but still no *de facto* solution. In particular, ontologies appeared to solve *semantic* interoperability in S16 and S18; however, just considering this type seems insufficient due to the need to integrate several IoT application layers separately from data acquisition and hardware/software platforms. Hence, three new types have been considered [70]: (i) *device* (to provide information exchange between physical and software components); (ii) *network* (to deal with seamless communication of devices over different networks); and (iii) *platform* (to offer a collaboration of the diverse platforms used in IoT due to diverse operating systems, programming languages, and access mechanisms for data and things). Additionally, *data* interoperability refers to a larger type that encompasses these three types to treat heterogeneity jointly. According to S18, interoperability standards must be sought to obtain the most cost-effective solutions in this scenario. Other more comprehensive and integrative solutions are Industry 4.0 reference architectures [65], including IIRA (Industrial Internet Reference Architecture) [44] and RAMI 4.0 (Reference Architecture Model for Industry 4.0) [45], the most relevant ones in the Industry 4.0 community. As a closer domain of Industry 4.0, **manufacturing** also comprises software systems that concern industrial product development and management life cycle. Companies use different strategies and technological sources across PDP (product development process), embodying different characteristics and perspectives to meet the customers' needs. Similar to Industry 4.0 domain, *semantic* interoperability is considered to solve the hardware and software heterogeneity (as pointed out in S13 and S25). Additionally, industry players apply meta-models, domain-specific models, and frameworks to cope with distinct perspectives (e.g., mechanical engineering, electrical engineering, and computer science) of information sharing across different phases of product development.

**Cyber-physical systems** have also brought challenges due to the need to integrate a heterogeneity of elements, including physical and software components, while they sometimes operate in critical application domains. Hence, beyond *technical*, *syntactic*, and *semantic* interoperability, other types appeared (*data*, *constructive*, *information*, *operational*, *programmatic*, and *systems*) to deal with specific aspects of these systems, such as equipment maintenance, acquisition of management information, and reliable information exchange (S8 and S14). In the same direction, the domain of **constructive systems** (i.e., those that manage the planning, design, construction, operation, and maintenance of buildings) also has its specificities and the need to interoperate different kinds of systems (such as for digital representations of physical and functional characteristics of places), leading the need of *business*, *data*, *process*, and *service* interoperability. Domain-specific models are solutions for those interoperability types (S19).

**Context-awareness software systems** use context information to characterize a given situation, person, object, or place. S17 presented frameworks to deal with *organizational* interoperability and overcome distinct context data representation. Surprisingly, *pragmatic* interoperability, an LCIM interoperability level that considers context information, does not appear in the study. Otherwise, **collaborative systems** focuses on *pragmatic* interoperability to achieve the desired interoperation since the context is taken into account in the message exchange. To cope with this interoperability type, meta-model, ontologies, and platforms have been used to manage the context, intention, and effect of messages exchange (S10).

A couple of studies addressed cloud computing, IoT, and blockchain. In particular, **cloud computing** is a paradigm managing a pool of virtualized resources at infrastructure, platform, and software levels to deliver them as services over the Internet, storing and managing large amounts of heterogeneous data [80]. Cloud providers are categorized into proprietary and open-source platforms, and most of them have their API, protocol, and data format [58]. Due to the lack of adoption of standards in such a heterogeneous environment, when users want to shift to another provider or use additional services from different providers, they face the lock-in situation (S9 and S11). As the most obvious solution for cloud interoperability seems to be standards, several organizations are working on proposals, such as ISO/IEC 17788:2014, DCRM (Distributed Computing Reference Model), OCCI (Open Cloud Computing Interface) specifications, CIMI (Cloud Infrastructure Management Interface), and CDMI (Cloud Data Management Interface). Like Industry 4.0 and cyber-physical systems, cloud computing also investigates *semantic* interoperability solutions, such as OWL, SPARQL Protocol, and RDF Query Language. Open libraries, model-based, and open-service solutions are under investigation to manage inter-cloud interoperability issues.

**IoT** creates an ecosystem in which people and things interconnect among them through the Internet. The information exchange occurs anytime and anywhere, with devices and sensors monitoring people and the environment and generating a large volume of data that different applications can consume. The heterogeneity of IoT (e.g., hardware, network, software infrastructure, applications, and data acquisition protocols) often hampers the interoperability of IoT-based systems. Hence, these systems often operate in vertical silos and adopt *semantic* interoperability approaches, such as Semantic Web technologies (e.g., OWL and RDF) and SSN (Semantic Sensor Network) ontology (S22 and S23). At the same time, the adoption of common semantics (for instance, through ontology or shared schema) cannot be viable in some cases, leading to the need for new types of low-level interoperability, such as *device* and *platform* (S24). Additionally, cloud and edge platforms (used to store data) and Software Defined Network (SDN) (a network-level integration approach) are used (S28).

**Blockchain** is a distributed ledger system that allows reliable transactions among untrusted network participants. As S30 pointed out, various use cases have diverse requirements, requiring the adoption of blockchains with different capabilities. As a result, like in the IoT domain, blockchain systems exist and operate in silos, and people cannot reap the full benefits of blockchain technology (S34 and S35). Blockchain interoperability enables smart contract invocations, asset exchanges, and data verification, increasing flexibility, application migration, and scalability. However, achieving interoperability poses several challenges, including security and privacy (S35). According to S30, blockchain interoperability deals with three significant aspects: (i) interoperability between different blockchains; (ii) interoperability between decentralized applications (dApps) using the same blockchains; and (iii) interoperability between blockchain and other technologies such as enterprise systems. In this context, interoperability makes possible that different transaction between different chains can be: (i) Cross-Chain Transaction (CC-Tx), which belongs to the same blockchain system (homogeneous blockchain); and (ii) Cross-Blockchain Transaction (CB-Tx) between different blockchains (heterogeneous blockchain). Moreover, the solution is often classified into blockchain of blockchains (BoB), public and hybrid connectors. BoB is usually referred as a platform for developers to construct Cross-Chain dApps (CC-dApps) spread across multiple blockchains (e.g., Polkadot, Cosmos, and Ethereum). Connectors-based solutions coordinate the token-transferring process on distinct ledgers and serve as a translator between different ledger protocols (e.g., HTLC (public) and ILP (hybrid)) (S30). An alternative way to classify solutions is presented in S37: (i) chain, target to public blockchains (e.g., Loom, Rootstock, and Herdius); (ii) bridge, that is a connectors component across blockchains (e.g., BTC Relay and Peace Relay); and (iii) dApp-based interoperability solutions, target provide interoperability among applications and blockchains (e.g., Overledger and Plebeus).

In summary, we can say that most domains concern with *semantic* interoperability, which has ontologies and conceptual models as the most recurrent solutions (found in S3, S9, S13, S16, S18, S22, S23, S24, S25, S28, and S31). Additionally, we observe *technical* and *syntactic* interoperability have been sometimes solved in different domains with the adoption of standards. Although standards facilitate the development of computational solutions, particularly in IoT and cloud computing, industry stakeholders have not widely adopted them, often leading to lock-in problems. Similar to IoT and cloud computing, the blockchain domain has faced heterogeneity issues but the studies deal with a specific type — the blockchain interoperability type — to address their solutions. Other common solutions in diverse domains are platforms, API, reference architectures, and gateways that have also provided *technical*, *syntactic* and *semantic* interoperability solutions (as in S24 and S26). Furthermore, the primary solutions to manage *organizational* interoperability (including *business*, *enterprise*, and *process*) are frameworks and conceptual models (S2, S3, S14, S15, and S27).

Our intention with this RQ is not to be exhaustive but to bring a broader view of what the secondary studies discussed. The domains mentioned above may have many other interoperability solutions and types; besides, practically all other existing domains present interoperability issues requiring specific investigation.

## 5 DISCUSSIONS

Interoperability refers to a non-functional or quality requirement required in almost all existing SIS. In this scenario, this work offers a comprehensive view of various interoperability types, models, and frameworks, as well as how different domains and associated solutions have addressed interoperability. Interoperability has been investigated since software systems became distributed some decades ago and has attracted considerable attention from industry, academia, and standardization organizations.

Several significant initiatives aim to provide interoperability in various domains. These initiatives involve multiple stakeholders from different countries and sectors to achieve effective collaboration among SIS. For instance, EIF for enterprise, Open Connectivity Foundation (OCF)[19], which focuses on providing a common framework for IoT devices, European Blockchain Services Infrastructure (EBSI)[20] is another initiative that aims to promote the interoperability of blockchain-based services across Europe, FHIR standard, EU's eHealth Digital Service Infrastructure (eHDSI)[21], Interoperability Standards Advisor (ISA)[22] for health different systems and countries, to cite a few.

Comparing our work with the closest work [33], we identified 36 interoperability types, while that work presented 64. Analyzing the reasons, we observed that most types from [33] deal with low-level technical aspects (e.g., *communications*,

---

[19] https://openconnectivity.org/
[20] https://digital-strategy.ec.europa.eu/en/policies/european-blockchain-services-infrastructure
[21] https://digitalhealtheurope.eu/glossary/ehdsi/
[22] https://www.healthit.gov/isa/

*electronics*, *telecommunications*, and *plug-and-play*), which were replaced by *technical* and *syntactic* interoperability. Other types are very generic (e.g., *high-layer*, *lower-layer*, *higher-layer*, and *specification-level*) or solution-oriented (e.g., *object-oriented*, *procedure-oriented*, and *product-to-product*), so they were short-lived. We also observe that both works ([33] and ours) identified three interoperability types as those most recurrent: *organizational*, *semantic*, and *technical*; besides, some social-technical types also appeared in both studies, e.g., *cultural*, *organizational*, *enterprise*, and *process*. In some way, both works can reveal the evolution of interoperability conceptualization and stabilization of some interoperability types. While [33] identified interoperability types through a survey on interoperability measurement, our study is more extensive by considering interoperability types, conceptual models, frameworks, and some domains that addressed interoperability solutions. Moreover, this related work was published 15 years ago, and the technologies have suffered considerable advances; therefore, we believe an updated view of systems interoperability types is valuable.

As we can see, interoperability types often guide research and development. However, there is still much work to be done, with several open issues remaining. The following sections present our main findings and related open issues, urgent actions to be performed, potential research opportunities, and the threats to the validity of our work.

### 5.1 Findings and Open Issues

Following, we discuss the main findings of our work and identify open issues associated with each finding, making it possible for researchers and practitioners further inspiration, reflection, and investigation.

- **Interoperability is a field in continuous evolution:** Interoperability has been an important topic of interest for at least four decades. Since the first works on interoperability at the beginning of the 1980s when technical and syntactic interoperability started to be treated [51], going to the 1990s when middleware for interoperability drew attention [13] and to 2000s when the military domain highlighted the need of non-technical interoperability [82], the concept of interoperability has smoothly evolved over the years to keep up with the continuous evolution of hardware and software technologies. It has also evolved, encompassing not only technical interoperability but also social-technical ones. Hence, while technical interoperability overcomes barriers regarding heterogeneity (e.g., software and hardware infrastructure), social-technical ones cope with differences (e.g., cultural, ethical, languages, doctrines, and legal aspects). At the same time, this evolution has signaled that popular interoperability types (i.e., technical, syntactic, and semantic) have no longer fulfilled the current systems interoperability needs [42]. For example, blockchain, IoT, and cloud computing have required new and disruptive means to achieve interoperability.
  **Open issues:** With the inevitable emergence of new software and hardware technologies and even disruptive SIS, the field of systems interoperability will need to evolve continually. New possibilities of interactions in SIS and suitable interoperability types should be characterized, while existing types should be revisited considering the new scenarios.
- **Interoperability has gone beyond the software system's boundary:** The emergence of new application domains (e.g., smart-* systems, Health 4.0, and Industry 4.0) has brought several interoperability concerns and challenges. Software systems have mediated interactions among people, organizations, and things (e.g., equipment, devices, and cars), often replacing activities previously performed by human beings. Hence, interoperability is today not only a software need but also a concern outside the software systems' boundary [56]. Consequently, providing interoperability in systems also involves solving real-world interoperability, e.g., when systems from two countries with their specific legislation need to interoperate, legal interoperability should be treated (as in the case of EIF as a proposal for EU). Therefore, the interoperability field has included interoperability types out of the boundary of software systems themselves (e.g., cultural, legal, and knowledge) as a natural way to connect organizations, different cultures, countries, and regions that software systems have made possible.

**Open issues:**: A critical challenge is to perceive the limits of computational support for solving some social-technical interoperability types (e.g., *legal*, *cultural*, and *political*). As these types are related to issues beyond software boundaries, finding suitable interoperability solutions should involve different stakeholders besides those from the computing area.

· **Interoperability must be considered from different perspectives**: Nowadays, software systems are responsible for different interactions (involving people, organizations and their processes, and things). Interoperability has increasingly dealt with heterogeneity in software systems regarding different elements (e.g., platforms, languages, communication protocols, and so on) and also differences that systems must address (e.g., cultures from different regions and legal issues from different countries). Besides those interoperability types found in our work, others have recently appeared in the literature, such as *human* [39], *social* [18], *political* [67], *full* [57], and *trustworthy* [2]. Additionally, similar to other quality requirements, interoperability must be treated along with all systems life cycle — from requirements engineering to system modeling, architectural design, implementation, and maintenance — which must involve teams from computing and also from other areas. In this scenario, it is undeniable that different perspectives must be considered for achieving suitable interoperation in software systems.

**Open issues:** Interoperability should become a multidisciplinary research field, encompassing other knowledge areas, like social sciences, economy, and politics.

· **Interoperability refers to a consensus, but the panorama does not reflect that yet**: The research works indicate that interoperability has drawn considerable attention, aligned with the industry interest in solving systems interoperability. Basically, the interoperability principle is to establish a consensus to achieve effective collaboration among systems; however, studies show that finding a consensus even within specific application domains like health is not trivial [25]. New interoperability types have also continuously emerged with little or no investigation yet. We also observe it lacks a consensus in the research community and industry on those basic types, those specific for specific application domains or classes of systems (e.g., SoS), or those for a given technology like IoT or cloud computing [59, 70]. In turn, research initiatives seem not to talk to each other.

**Open issues:** A consensual understanding of the existing interoperability types is not still established. It is also required interoperability solutions, strategies to solve misunderstandings, consolidation of many interoperability types, and avoidance of new types without considering how existing ones should be a goal to pursue. Indeed, the interoperability concept still needs to be better and widely comprehended, including types considered as social-technical interoperability.

· **Interoperability models and frameworks are good solutions, but their hierarchical levels need to be rethought**: Interoperability models (i.e., conceptual models and IAM) and frameworks play an important role in organizing and sharing interoperability knowledge in a systematized way. Several areas recognize the value of interoperability models for systems interoperability, so these models have been used as a reference guide for industry solutions and research works. Several areas also recognize that frameworks like EIF and Athena can be very promising in specific domains. Nevertheless, sometimes these models and frameworks do not present a widespread adoption [23]. Furthermore, they do not consider new interoperability types that seem to be essential in systems that they support the development. For instance, INTEROP (a framework for enterprise systems) could address *legal* interoperability as nowadays enterprise systems have crossed the countries' borders. Most models and frameworks provide a hierarchical organization of the interoperability types, in which *technical*, *syntactic*, and *semantic* comprise a base stack for systems interoperation. At the same time, new types emerged to deal with the interactions among elements in a given system or interactions among various systems and may require different types that seem not suitable to be organized hierarchically. Some examples of these types are *legal*, *cultural*, and *political*.

**Open Issues:** As interoperability models and frameworks can no longer reflect the current interoperability scenarios, they need to be updated with the new interoperability types and new kinds of relationships than hierarchical one. Specifically regarding IAM, certifications of interoperability levels for systems could be considered.

- **Domains with success interoperability should be an example to be followed**: Internet/Web domains successfully adopt several standards [93] to solve *technical* and *syntactic* interoperability. Hence, web-based systems have adopted them and have solved other types (in particular, *semantic* and *organizational*). Furthermore, health stands out as one of the domains found in our study that has mostly investigated interoperability. The work of Mello et al. [60] better summarizes the main works and solutions for interoperability in health systems. This domain has already examined the impact of cloud computing and IoT heterogeneity on the interoperability of its systems. For a while, the blockchain domain seems to be a prime and recent example of efforts to address interoperability by presenting initiatives that facilitate the development of interoperability solutions, ultimately making blockchain technology more accessible and valuable to a broader range of industries. Solutions to decrease lock-in situations also exist and should be followed.

  **Open Issues**: As the interoperability solutions, mainly in highly heterogeneous scenarios, are costly and time-consuming to develop, success domains and their solutions (e.g., Internet/Web and health) should serve as an example for others and for new domains like Industry 4.0.

- **Several interoperability solutions exist but still without a broader adoption:** Several interoperability solutions, including standards, ontologies, frameworks, conceptual models, platforms, reference architectures, API, services, and gateways, have supported the achievement of systems interoperability over the years. In general, domains that present heterogeneity in their components (which need to interoperate) are the ones with a variety of solutions [42]. In particular, standards are a cost-effective solution but challenging to be adopted by stakeholders and players that already use their particular solutions. For instance, although several works exists in IoT domain, applications usually run in vertical silos [70]. In general, the same happens with other solutions that are usually isolated contributions without a wide dissemination even in their domains.

  **Open issue**: To make possible wide adoption of interoperability solutions, consortia encompassing companies, research institutions, standardization organizations, and even governments should be formed. For instance, Autosar[23] is a large consortium leveraging the adoption of a reference architecture. Moreover, interoperability solutions for systems involving newer technologies, such as edge, fog, blockchain, and artificial intelligence, are also required and could reuse success experiences from previous solutions, but disruptive ones will be possibly necessary.

- **A complete body of knowledge on systems interoperability is missing:** Our study presents an initial body of knowledge of systems interoperability; however, it also reveals that this knowledge is still somehow fragmented, even within specific application domains or technologies such as IoT. Moreover, several studies reported a lack of comprehensive information about interoperability types [8, 33, 36] or seek definitions of interoperability [63, 85]. Regarding interoperability models (including conceptual models and IAM) and frameworks, it lacks a consensus on what precisely they should comprise to be useful [75]. In this scenario, our work provides a step towards a holistic view of systems interoperability.

---

[23]https://www.autosar.org/

**Open Issue:** The knowledge about systems interoperability could be stated through a document similar to SEBOK (Systems Engineering Body of Knowledge) [27] or SWEBOK (Software Engineering Body of Knowledge) [15]. This document could specify the accepted definitions of interoperability types, their relationships, and what interoperability models and frameworks are exactly.

### 5.2  Future Actions and Potential Research Opportunities

The interoperability types have been the driving force of much of the research work in the field. In this work, the analysis of various interoperability types and associated topics, such as models and frameworks, revealed several open issues, as discussed previously. In response, we present some of the most urgent and future actions and potential research opportunities that could contribute to advancing the state of the art in the field.

- **Interoperability discussion forums**: Interoperability has been discussed in several forums (like conferences and workshops), but as a subtopic associated with others such as in the architectural design of cyber-physical systems or health systems. We highlight the need of creating dedicated forums for specifically discussing interoperability challenges and solutions, bringing together researchers and practitioners from various fields, including computer science, engineering, politics, and social sciences. Such forums could bring a much-needed common understanding on systems interoperability and how to achieve and ensure it, as well as serve as a venue for sharing ideas, best practices, and case studies and address both theoretical and practical interoperability aspects. These forums could also focus on specific topics, such as data interoperability, social interoperability, full interoperability, technical standards, and multidisciplinary aspects such as the implications of legal interoperability.
- **Interoperability task forces and standardization organization**: Following successful examples, such as the OMG (Object Management Group), W3C (World Wide Web Consortium), IEEE (Institute of Electrical and Electronics Engineers), and HIMSS (Healthcare Information Management Systems Society), we suggest to create task forces and an organization specifically focused on systems interoperability. These task forces could be organized around interoperability challenges like in cloud computing, IoT, blockchain, smart cities, or financial systems. Moreover, such organization could address: (i) the development of interoperability patterns (or interoperability standards as those existing in some domains), guidelines, and solutions that can be shared and reused across different domains; and (ii) the establishment of interoperability certification programs that could promote best practices. Finally, this organization could be a vehicle to become transparent what has been done in different domains, promoting the sharing of their achievements.
- **SIS Interoperability Body of Knowledge (SISBOK)**: We foresee the need of a comprehensive and evolving body of knowledge of SIS Interoperability (SISBOK). SISBOK would be a repository of knowledge related to interoperability, including types, models, frameworks, and related solutions, serving as a basis for continuous evolution of these concepts and their definitions, as a reference for researchers and practitioners, and as a learning object for students. The development of SISBOK will require involvement of multiple stakeholders, including researchers, practitioners, and standardization organizations. Knowledge from previous experience, e.g., from existing interoperability models, should be also considered and incorporated in SISBOK when pertinent, although most models require updated.
- **Detailed definition of the existing interoperability types**: A deeper understanding of the interoperability types can drive the progress in the field. A detailed, clear, and common definition of each interoperability type considering heterogeneity and differences among them, as well as requirements, constraints, and attributes, can enhance stakeholders' communication. Interoperability frameworks and patterns could also adopt unified definitions for their types. Additionally, meta-frameworks could provide a high-level understanding of the relationships among different types.

- **Investigation of social-technical interoperability**: Several social and organizational factors, such as people's culture, ethics, politics, and culture and structures of organizations, directly impact the achievement or not of different levels of SIS interoperability. These factors have not been sufficiently investigated, so requiring a deeper examination of the various scenarios in which social-technical interoperability should be treated, the differences influencing these factors, and strategies to overcome such differences. In addition, investigating the relations between technological and social-technical interoperability types in different contexts (including new types and context where they will likely emerge) could leverage the SIS interoperability.

- **New interoperability types:** Identifying and defining new interoperability types are essential to the development and evolution of SIS, especially considering large and complex SIS that depend on diverse new technologies like blockchain. For instance, new scenarios where SIS will operate have given increasing importance to ethical issues and data protection and privacy, leading to the need for new interoperability types, which should also be consistent with types already in use.

- **Remodeling the relationships among interoperability types**: Understanding the relationships among types is also crucial to achieve SIS interoperability. Considering our investigation presented in this work, we observe that it is necessary to remodel the relationships among those types, going beyond the well-known hierarchical relationships. For instance, legal interoperability has a cross-cutting nature, affecting both technological and social-technical perspectives of an SIS. By understanding how different types relate each other, stakeholders can make more informed decisions of the required types for a given SIS. Additionally, interoperability models, frameworks, patterns, and solutions should be updated accordingly.

- **Assessment models for interoperability solutions:** Existing IAM provide a valuable support for evaluating and selecting interoperability solutions, but they are somehow outdated considering the very new interoperability solutions emerging continuously. Moreover, currently, it is unclear which interoperability types each solution offer. Hence, new criteria and methods to evaluate these solutions and even new assessment models are required. Additionally, IAM could designate "stamps" indicating the interoperability types covered by the solutions, so helping stakeholders in selecting more adequate solutions.

- **Maturity model for SIS interoperability:** Maturity models to assess SIS regarding their interoperability levels could be another potential future research. These models should consider different types, systems requirements, and possibly particularities of specific domains. In addition, they should outline the steps needed for a given SIS to reach each maturity level and progress to higher levels.

## 5.3 Threats to Validity

We identified a set of threats to the validity for our tertiary study and the countermeasures to mitigate them, according to the work of Ampatzoglou et al. [5]:

- **Study Selection Validity:** The first threat was relevant studies that might have been missed; hence, we adopted several countermeasures to deal with this threat. In particular, the search string might have imposed threats during the studies search. To mitigate it, we followed the process presented in [50] to test, calibrate, and define our string carefully. Another countermeasure was the search for studies in three publication databases, namely Scopus, IEEE Xplore, and ACM DL, since they are considered the most relevant sources of studies in software engineering [29, 50].

During the search, we also considered various types of scientific publications, including conference papers, journal articles, technical reports, books, and book chapters, as a countermeasure to enlarge the scope of our selection.

Additionally, to ensure an unbiased selection of studies, we defined three RQ and eight selection criteria (two inclusion criteria and six exclusion criteria) in advance. We believe the RQ (and associated rationale) and the criteria were detailed enough to provide confidence in how we obtained the final set of studies. Moreover, aiming to increase the reliability of selection, at least two authors checked and read each study as a way to mitigate the threat. When conflicts in the selection criteria application occurred, the third and fourth authors reviewed the study to make the final decision. Another countermeasure was that all authors (including two with extensive experience conducting secondary studies) reviewed the tertiary study protocol; additionally, we systematically and rigorously followed the protocol to avoid bias and ensure the inclusion of all relevant studies. However, some studies could have been excluded due to the lack of information in the title, abstract, keywords, introduction, and conclusion sections; furthermore, studies could have been missed despite all our effort to include all relevant studies. After applying these countermeasures, we believe the final set of 37 studies reflects the state of the art in SIS interoperability.

- **Data validity:** Another threat was that we might have missed relevant data from the secondary studies. Hence, a countermeasure was using a data extraction form that was carefully prepared and checked to make it possible to collect all data from the studies and answer our RQ correctly. We also performed a pilot test with the data extraction form to observe its suitability and efficiency in gathering data.

  Additionally, as some data from secondary studies was not apparent to be easily extracted, we had to interpret it. As a countermeasure, we had a specialist (one of the authors) who doubled checked each data extracted and the interpretations done. This specialist identified some inconsistencies and incompatibilities that we discussed until reaching a consensus. We also used the extracted data to trigger several discussion sessions to understand the data better and identify data correlation.

- **Research validity:** It refers to the possibility that our opinion or knowledge might have influenced the synthesis of the results. As a countermeasure, all authors of this study contributed to discussing and synthesizing the results. In turn, the authors have researched systems interoperability and software architecture for years; hence, we believe our previous knowledge could have contributed to a better synthesis of results.

  We also previously defined and systematically followed a detailed protocol (as summarized in Section 3.1), which was based on well-established guidelines for tertiary studies [50]. In addition to the information presented in this work, we offer external material with all raw and extracted data on which we based our findings, as well as information to make possible the reproducibility and auditability of this tertiary study.

## 6 CONCLUSIONS

Implementing interoperability in SIS is very challenging not only in large technology-oriented systems with complicated interactions but also in small and medium-sized systems where a balance of the benefits and drawbacks of adopting existing interoperability models, frameworks, or solutions is required. Organizations, research communities, and industries demonstrate a continuous interest in better understanding the different interoperability types and the actual relationships among them and in proposing different solutions.

In this scenario, our tertiary study systematically retrieved 586 studies (494 during the first conduction and 92 during the update) and scrutinized 37 studies that satisfied our criteria and were published from 2012 to 2023. This work contributes with an updated and broad view of SIS interoperability, including 36 interoperability types, distilling their definitions and classifying them, and different ways to organize them that several models and frameworks have provided. We observe these types that emerged over the years reflect how the field has evolved, showing the trends and serving as a basis for solutions. Considering such types, we analyzed several models, frameworks, and diverse domains concerned with interoperability. We identified important findings, open issues, urgent actions to be further performed, and potential research opportunities. Therefore, this work can be considered a step toward a body of knowledge on SIS interoperability.

For the future, a critical step is to change the mindset of practitioners and researchers that interoperability must be treated as a multidisciplinary field, which should also deal with interoperability outside the systems (e.g., legal, political, and cultural interoperability). In this perspective, this work intends to call attention to the current scenario of the field, increase the awareness of SIS interoperability, and contribute with a body of knowledge that should be updated continuously to keep it aligned with the state of the art in the field, which is indeed dynamic.

**Acknowledgments:** This research was funded by Bahia Research Foundation - FAPESB (Grant: TIC0002/2015), São Paulo Research Foundation - FAPESP (Grants: 2022/03276-6, 2015/24144-7), and National Council for Scientific and Technological Development - CNPq (Grant: 313245/2021-5).

30	Maciel, R.S.P., et al.[18] Bryant, R., Dortmund, A., Lavoie, B., 2020. Social interoperability in research support: Cross-campus partnerships and the university research enterprise. OCLC Online Computer Library Center .

[19] Burns, T., Cosgrove, J., Doyle, F., 2019. A review of interoperability standards for industry 4.0. Procedia Manufacturing 38, 646–653.

[20] Burzlaff, F., Wilken, N., Bartelt, C., Stuckenschmidt, H., 2019 (2022). Semantic interoperability methods for smart service systems: A survey. IEEE Transactions on Engineering Management 69, 4052–4066.

[21] C4ISR Architecture Working Group and others, 1998. Levels of information systems interoperability (LISI). United States of America Department of Defense .

[22] C4ISR Interoperability Working Group, 1998. Levels of information systems interoperability (LISI). Technical Report. US Department of Defense, Washington, DC.

[23] Casiano Flores, C., Rodriguez Müller, A.P., Albrecht, V., Crompvoets, J., Steen, T., Tambouris, E., 2021. Towards the inclusion of co-creation in the european interoperability framework, in: International Conference on Theory and Practice of Electronic Governance, pp. 538–540.

[24] Chen, D., 2006. Enterprise interoperability framework, in: EMOI-INTEROP.

[25] Chetty, M., Botha, A., Herselman, M., 2020. An instantiation of a process model towards health interoperability, in: South African Institute of Computer Scientists and Information Technologists 2020, pp. 180–188.

[26] Clark, T., Jones, R., 1999. Organisational interoperability maturity model for c2, in: Proceedings of the 1999 Command and Control Research and Technology Symposium, Citeseer.

[27] Cloutier, R.J., 2017. Guide to the Systems Engineering Body of Knowledge (SEBoK), version 1.8. URL: https://www.sebokwiki.org. (Accessed in 10/03/2023).

[28] Dersin, P., 2014. Systems of systems. IEEE-Reliability Society. Technical Committee on "Systems of Systems", url = https://rs.ieee.org/ technical-activities/technical-committees/ systems-of-systems.html, note = (Accessed in 10/03/2023),.

[29] Dyba, T., Kitchenham, B.A., Jorgensen, M., 2005. Evidence-based software engineering for practitioners. IEEE Software 22, 58–65.

[30] EIF, 2017. Interoperability solutions for public administrations, businesses and citizens. About ISA2. Technical Report. https://ec.europa.eu/isa2/isa2_en.

[31] Elheni-Daldoul, D., Le Duigou, J., Eynard, B., Hajri-Gabouj, S., 2013. Enterprise information systems' interoperability: Focus on plm challenges, in: IFIP International Conference on Advances in Production Management Systems (APMS), pp. 184–191.

[32] Ezzat, S.K., Saleh, Y.N., Abdel-Hamid, A.A., 2022. Blockchain oracles: State-of-the-art and research directions. IEEE Access 10, 67551–675772.

[33] Ford, T., Colombi, J., Graham, S., Jacques, D., 2007. Survey on interoperability measurement. Technical Report. Air Force Institute of Technology, US.

[34] Fraga, A.L., Vegetti, M., Leone, H.P., 2020. Ontology-based solutions for interoperability among product lifecycle management systems: A systematic literature review. Journal of Industrial Information Integration 20, 100176.

[35] Guédria, W., Chen, D., Naudet, Y., 2009. A maturity model for enterprise interoperability, in: OTM Confederated International Conferences" On the Move to Meaningful Internet Systems", Springer. pp. 216–225.

[36] Gürdür, D., Asplund, F., 2018. A systematic review to merge discourses: Interoperability, integration and cyber-physical systems. Journal of Industrial information integration 9, 14–23.

[37] Gürdür, D., Asplund, F., El-khoury, J., 2016. Measuring tool chain interoperability in cyber-physical systems, in: 11th System of Systems Engineering Conference (SoSE), pp. 1–4.

[38] Guédria, W., Gaaloul, K., Proper, H., Naudet, Y., 2013. Research methodology for enterprise interoperability architecture approach, in: International Conference on Advanced Information Systems Engineering Workshops (CAiSE), pp. 16–19.

[39] Handley, H., 2014. A network model for human interoperability. Human Factors 56, 349–360.

[40] Hanssen, G., Šmite, D., Moe, N., 2011. Signs of agile trends in global software engineering research: A tertiary study, in: IEEE 6th International Conference on Global Software Engineering Workshop (ICGSE-W), pp. 17–23.

[41] Haugum, T., Hoff, B., Alsadi, M., Li, J., 2022. Security and privacy challenges in blockchain interoperability-a multivocal literature review, in: Proceedings of the International Conference on Evaluation and Assessment in Software Engineering 2022, pp. 347–356.

[42] Hazra, A., Adhikari, M., Amgoth, T., Srirama, S.N., 2021. A comprehensive survey on interoperability for iiot: taxonomy, standards, and future directions. ACM Computing Surveys (CSUR) 55, 1–35.

[43] IEEE, 1990. IEEE standard glossary of software engineering terminology (IEEE Std 610.12-1990). CA: IEEE Computer Society 169.

[44] Industrial Internet Consortium, 2019. Industrial Internet Reference Architecture (IIRA). URL: https://www.iiconsortium.org/IIRA.htm. (Accessed in 10/03/2023).

[45] International Electrotechnical Commission, 2017. IEC PAS 63088:2017 Smart manufacturing - Reference architecture model Industry 4.0 (RAMI4.0).

[46] Jiang, S., Jakobsen, K., Jaccheri, L., Li, J., 2021. Blockchain and sustainability: A tertiary study, in: IEEE/ACM International Workshop on Body of Knowledge for Software Sustainability (BoKSS), pp. 7–8.

[47] Kaur, K., Sharma, D.S., Kahlon, D.K.S., 2017. Interoperability and portability approaches in inter-connected clouds: A review. ACM Computing Surveys (CSUR) 50, 1–40.

[48] Kingston, G., Fewell, S., Richer, W., 2005. An organisational interoperability agility model. Technical Report. Defence Science and Technology Organisation Canberra (Australia).

**A. Structure of Interoperability Models and Frameworks**

Table 5. Interoperability types (or level) of interoperability models and frameworks

| Conceptual Model | Structure |
|---|---|
| SoIM | Level 1: Separate systems, Level 2: Shared resources, Level 3: Gateways, Level 4: Multiple entry points, Level 5: Conformable/compatible systems, Level 6: Completely interoperable systems |
| MCISI | Full interoperable systems, Partial interoperable systems, Non-interoperable systems |
| LCI | Technical (physical, protocol, data/object model, and information), Organizational (aligned procedures, aligned operations, harmonized, strategies/doctrines, and political objectives) |
| NMI | Degree 1: Unstructured data exchange, Degree 2: Structured data exchange, Degree 3: Seamless sharing of data, Degree 4: Seamless sharing of information |
| LCIM | Conceptual, Dynamic, Pragmatic, Semantic, Syntactic, Technical |
| SoSI | Technical, Programmatic, Constructive Operational |
| NMI | Degree 0: No data exchange, Degree 1: Unstructured data exchange, Degree 2: Structured data exchange, Degree 3: Seamless sharing of data, Degree 4: Seamless sharing of information |
| **Interoperability Assessment Model** | **Structure** |
| LISI | Level 0: Isolated - manual, Level 1: Connected - peer-to-peer, Level 2: Functional - distributed, Level 3: Domain - integrated, Level 4: Enterprise - universal |
| OIM | Independent, Cooperative, Collaborative, Combined, Unified |
| OIAM | Level 0: Static, Level 1: Amenable, Level 2: Accommodating, Level 3: Open, Level 4: Dynamic |
| EIMM | Enterprise, Business, Organizational, Service, Legal |
| MMEI | Business, Process, Service, Data |
| ULSSIMM | Technical, Syntactic, Semantic, Organizational |
| DIAM | Business, Process, Service, Data |
| **Interoperability Framework** | **Structure** |
| AIF | Enterprise, Business, Process, Service, Data |
| IDEAS | Application, Business, Communication, Data, Knowledge |
| EIF | Legal, Organizational, Semantic, Technical |
| GridWise | Organizational, Informational, Technical |
| INTEROP | Business, Data, Process, Service |
| BIF | Public connectors, Blockchain of blockchain, Hybrid connectors (Reffered to as categories) |